\documentclass[fleqn,usenatbib]{mnras}

\usepackage[T1]{fontenc}
\usepackage{graphicx}
\usepackage{amsmath}
\usepackage{amssymb}
\usepackage{longtable}
\usepackage{rotating}
\usepackage{lscape}
\usepackage{hyperref}
\usepackage{newtxtext,newtxmath}

\newcommand{\logNHI}{\ensuremath{\log N(\mbox{\ion{H}{i}})}}

\def\hi{H~{\sc i}}
\newcommand{\nhi}{\ensuremath{N(\mbox{\ion{H}{i}})}}
\def\mgii{Mg~{\sc ii}}

\title[MUSE-ALMA Haloes Overview]{MUSE-ALMA Haloes VII: Survey Science Goals \& Design, Data Processing and Final Catalogues}

\author[P\'eroux et al.]{C. P\'eroux$^{1,2}$\thanks{E-mail: celine.peroux@gmail.com}, S. Weng$^{1,3,4,5}$, A. Karki$^{6}$, R. Augustin$^{7}$, V. P. Kulkarni$^{6}$, R. Szakacs$^{1}$,
\newauthor
A. Klitsch$^{8}$, A. Hamanowicz$^{7}$, A. Y. Fresco$^{9}$, 
M. A. Zwaan$^{1}$, A. Biggs$^{1}$, A. J. Fox$^{10}$, 
\newauthor
M. Hayes$^{11}$, J. C. Howk$^{12}$, G. G. Kacprzak$^{13, 5}$, S. Kassin$^{7}$, H. Kuntschner$^{1}$, D. Nelson$^{14}$ 
\newauthor
\& M. Pettini$^{15}$
\\
$^{1}$ European Southern Observatory, Karl-Schwarzschildstrasse 2, D-85748 Garching bei M{\"u}nchen, Germany\\
$^{2}$ Aix Marseille Universit\'e, CNRS, LAM (Laboratoire d'Astrophysique de Marseille) UMR 7326, 13388, Marseille, France \\
$^3$ Sydney Institute for Astronomy, School of Physics, University of Sydney, NSW 2006, Australia\\
$^4$ ATNF, CSIRO Astronomy and Space Science,  PO Box 76, Epping, NSW 1710, Australia \\
$^5$ ARC Centre of Excellence for All Sky Astrophysics in 3 Dimensions (ASTRO 3D)\\ 
$^6$ Department of Physics and Astronomy, University of South Carolina, Columbia, SC 29208, USA\\
$^{7}$ Space Telescope Science Institute, 3700 San Martin Drive, Baltimore, MD 21218, USA \\
$^{8}$ DARK, Niels Bohr Institute, University of Copenhagen, Jagtvej 128, 2200 Copenhagen, Denmark\\
$^{9}$ Max-Planck-Institut f\"ur Extraterrestrische Physik (MPE), Giessenbachstrasse 1, D--85748 Garching, Germany\\
$^{10}$ AURA for ESA, Space Telescope Science Institute, 3700 San Martin Drive, Baltimore, MD 21218\\
$^{11}$ Stockholm University, Department of Astronomy and Oskar Klein Centre for Cosmoparticle Physics, AlbaNova University Centre, SE-10691, Stockholm, Sweden\\
$^{12}$ Department of Physics, University of Notre Dame, Notre Dame, Indiana 46556, USA\\
$^{13}$ Centre for Astrophysics and Supercomputing, Swinburne University of Technology, Hawthorn, Victoria 3122, Australia\\
$^{14}$ Universit\"at Heidelberg, Zentrum f{\"u}r Astronomie, Institut f\"ur theoretische Astrophysik, Albert-Ueberle-Str. 2, 69120 Heidelberg, Germany\\
$^{15}$ Institute of Astronomy, University of Cambridge, Madingley Road, Cambridge CB3 0HA, UK\\
}

\date{}
\pubyear{2021}

\begin{document}
\maketitle

\begin{abstract}
The gas cycling in the circumgalactic regions of galaxies is known to be multi-phase. The MUSE-ALMA Haloes survey gathers a large multi-wavelength observational sample of absorption and emission data with the goal to significantly advance our understanding of the physical properties of such CGM gas. A key component of the MUSE-ALMA Haloes survey is the multi-facility observational campaign
conducted with VLT/MUSE, ALMA and HST. MUSE-ALMA Haloes targets comprise 19 VLT/MUSE IFS quasar fields, including 32 $z_{\rm abs}<$0.85 strong absorbers with measured \nhi\ $\geq 10^{18}$ cm$^{\rm -2}$ from UV-spectroscopy. We additionally use a new complementary HST medium program to characterise the stellar content of the galaxies through a 40-orbit three-band UVIS and IR WFC3 imaging. Beyond the absorber-selected targets, we detect 3658 sources all fields combined, including 703 objects with spectroscopic redshifts. This galaxy-selected sample constitutes the main focus of the current paper. We have secured millimeter ALMA observations of some of the fields to probe the molecular gas properties of these objects. 
Here, we present the overall survey science goals, target selection, observational strategy, data processing and source identification of the full sample. Furthermore, we provide catalogues of magnitude measurements for all objects detected in VLT/MUSE, ALMA and HST broad-band images and associated spectroscopic redshifts derived from VLT/MUSE observations. Together, this data set provides robust characterisation of the neutral atomic gas, molecular gas and stars in the same objects resulting in the baryon census of condensed matter in complex galaxy structures.

\end{abstract}



\section{Introduction}

Only a minority of the normal matter in the Universe can be probed by observations of starlight from galaxies. The remaining 90~per~cent of the baryons reside in the so-called interstellar and intergalactic gas. The temporal and spatial evolution of these baryons is best traced by studies of the physical processes by which gas travels into, through, and out of galaxies. The sites of these gas exchanges are the immediate surroundings of galaxies, the so-called circum-galactic medium or CGM \citep[e.g.][]{shull14, Tumlinson17}. On these galactic scales, we refer to the combination of stars and neutral (atomic plus molecular) gas as condensed matter. More globally, the cosmic baryon cycle describes these processes of motion and transformation of the baryons \citep[e.g.][]{peroux2020}.  

The canonical picture has galaxy growth being fed by inflows of gas from the intergalactic medium, IGM \citep[e.g.][]{Dekel09}. These baryons from the cosmic web cool into a dense atomic then a molecular phase, which fuels star formation. Once stars are formed, galaxies enrich the IGM with ionising photons and heavy elements formed in stars and supernovae, by driving galactic and AGN-driven winds into the CGM \citep{pettini2003}, some of which will fall back onto the galaxies in so-called galactic fountains \citep{Shapiro76, Fraternali17, Bish19}. A detailed probe of gas inflows and outflows is of
paramount importance for understanding these processes. 
Since gas, stars, and metals are intimately connected, gas flows affect the history of star formation and chemical enrichment in galaxies. Therefore the study of the multi-phase (cool T$<10^4$K, warm $10^4<$T$<10^5$K and hot $>10^5$K) CGM in particular is crucial for understanding the conversion of gas into stars.
Years of deep spectroscopic surveys of
galaxies \citep[e.g.][]{bordoloi2011, steidel2010}, absorption line studies in quasar spectra \citep[e.g.][]{turner17, Chen21} and sophisticated numerical simulations \citep[e.g.][]{keres12, schaye15, nelson15b, dave2017} have placed
interactions between galaxies and the CGM at the centre of our quest to understand the formation of galaxies and the growth of structure. 
However, determining what drives the physical
processes at play in the CGM still remains a complex problem in galaxy formation, in large part due to the lack of significant observational constraints.

At present, direct detection of the CGM in emission poses an observational challenge due to its diffuse nature (with hydrogen densities of the order of $n_{\rm H}<$0.1 cm$^{-3}$). Cosmological hydrodynamical simulations concur that the emission signal is faint by current observational standards \citep{augustin2019, peroux2019, corlies20, Wijers20, Wijers21, byrohl2021, Nelson21}. For these reasons, detections in emission at high-redshifts are currently limited to deep fields \citep{wisotzki2016, leclercq2017, wisotzki2018, leclercq2020, leclercq22} or regions around bright quasars \citep{cantalupo2005, arrigoni-battaia2015, Farina19, Lusso19, Mackenzie21} {while detections at z$<$1 are now becoming available \citep{epinat18, johnson18, Chen2019, rupke19, helton21, burchett21, zabl21}.}
Given this limitation, absorption lines detected against bright background quasars at UV and optical wavelengths provide the most compelling way to study the distribution, kinematics and chemical properties of CGM atomic gas to date. In these quasar absorbers, the minimum column density \citep[which is tightly correlated to the volumic gas density, see][]{rahmati13} that can be detected is set by the {apparent} brightness of the background sources and thus the detection efficiency is independent of the redshift of the foreground absorber host galaxy. In addition, absorption line-based metallicity measurements are independent of excitation conditions \citep{kewley2019, maiolino2019}. In fact, unlike emission lines metallicity estimates, they are largely insensitive to density or temperature and {high column density systems tracing neutral gas} require no assumption on a local source of excitation \citep{vladilo01, dessauges03}. 
Importantly, {multiple state-of-the-art cosmological hydrodynamical simulations and early observational results indicate that} the chemical properties of the CGM gas probed in absorption show an inhomogeneous metal distribution around galaxies with indication of a trend with galaxy orientation \citep{peroux2020, Wendt21}.

Recent technological advances related to 3D Integral Field Spectroscopy (IFS), which produces data cubes where each pixel on the image has a spectrum, have opened a new window for examining the CGM gas. This approach combines the information gathered in absorption against background sources (whose lines of sight pass through a galaxy's CGM) with traditional emission-based properties of galaxies. Following at least two decades of limited success in identifying the galaxies associated with quasar absorbers, IFS have open a new era in establishing the relation between absorption and emission with high success rates. Early efforts with near-infrared IFS VLT/SINFONI \citep{bouche2007, peroux2011, peroux2013, peroux2016} led to efficient discoveries of star-forming galaxies associated with \mgii\ and \hi\ absorbers at $z\sim$2 \citep[see also][]{rudie2017, Joshi21}. The optical IFS VLT/MUSE \citep{bacon2010} has proved to be a true game-changer in the field. Early on, the MUSE Guaranteed Time Observations (GTO) team established surveys including MUSE-QuBES \citep{Muzahid20} and MEGAFLOW \citep{schroetter2016, bouche2016, zabl2019, schroetter2019, Zabl2021} to relate gas traced by absorbers to galaxies. In a parallel effort, the MAGG survey targets higher redshift galaxies \citep{fumagalli2016, lofthouse2020, Dutta20}. {The Cosmic Ultraviolet Baryon Survey CUBS instead is absorption-blind and uncovers new quasar absorbers in a wide range of column densities (ranging from few times 16.0 $<$ \logNHI $<$20.1) at z$<$1 \citep{chen20, boettcher21, zahedy21, cooper21}.} 
By extending to bluer wavelengths, the optical IFS Keck/KCWI \citep{martin2010} has enabled similar studies at higher spectral resolution \citep{martin2019, Nielsen20}. BlueMUSE, a blue-optimised, medium spectral resolution IFS based on the MUSE concept and proposed for the Very Large Telescope is also under planning \citep{richard2019}. Contemporary to these works, ALMA - which can be viewed as an IFS at mm-wavelengths - has enabled the detections of both CO and [CII] emission {in galaxies} associated with strong quasar absorbers at intermediate and high redshifts, respectively \citep{neeleman2016, Klitsch18, neeleman2018, kanekar2018, neeleman2019, peroux2019, Klitsch21, Szakacs21}. These lines enable us to trace the colder ($\sim$100K) and denser phase of the neutral gas: the molecular hydrogen, H$_2$. The molecular gas constitutes the ultimate phase of the gas reservoir from which stars form and hence is an essential link to the baryon cycle. Together, these IFS observations have provided unique information on the resolved galaxy kinematics which can then be combined with the gas dynamics to probe gas flows in the CGM regions \citep{bouche2013, rahmani2018a, schroetter2019, zabl2019, Neeleman20, Szakacs21}.






Building on these successes, the MUSE-ALMA Haloes\footnote{\url{https://www.eso.org/\~cperoux/MUSE_ALMA_Haloes.html}} project aims at quantifying the physical properties of the gaseous haloes of galaxies with a particular focus on the multi-phase nature of the baryons in the CGM. To this end, the survey combines multi-facility campaigns which together gather information on the atomic, molecular and ionised gas as well as stellar populations (collectively refered to as condensed matter) in a sample of galaxies. The overarching objectives of the MUSE-ALMA Haloes survey are: i) to locate of gas with respect to galaxies whose stellar properties are also well established; ii) to characterise the amount and distribution of metals in the interstellar medium of galaxies as well as in the CGM gas; iii) to establish the dynamics of the gas in all its condensed forms; and iv) to perform the global census of baryons in galaxies and their CGM haloes. The initial study focussed on six quasar fields 
followed-up with ALMA. Early findings indicate that a fraction of the gas probed in absorption is likely related to intragroup gas \citep{peroux2017}. \cite{rahmani2018a} perform a detailed component-by-component analysis and report evidences for accreting gas onto a warped disk. Other quasar absorbers are tracing outflows with velocities such that the gas will remain bound to the host galaxy \citep{rahmani2018b}. By combining MUSE and ALMA data of the same quasar absorber for the first time, \cite{Klitsch18} offer a combined study of the kinematics of the neutral, molecular, and ionised gas. \cite{peroux2019} further use similar information together with dedicated hydrodynamical cosmological simulations to infer that a large fraction of the absorbing gas is likely the signature of gas with low surface brightness. \cite{hamanowicz2020} combine the data available to date and derive a high success rate in detecting galaxies associated with absorbers (89~per~cent). The authors find that most absorption systems are associated with pairs or groups of galaxies. Finally, \cite{Szakacs21} use new ALMA data to show that ionised and molecular gas phases within the disk are strongly coupled. The authors also report a new case of inflowing gas inferred from detailed kinematic study. 

Taken together, these results show an occasional mismatch in phase space of rotating disk kinematics and absorption profiles. These findings add to the paradigm shift where our former view of strong-\nhi\ quasar absorbers being associated to a single bright galaxy changes towards a picture where the \hi\ gas probed in absorption is related to complex galaxy structures associated with small groups or filaments. 
The conclusions also demonstrate that our understanding of the
physical properties of the CGM of complex group environments will benefit from associating the kinematics of individual absorbing components with each specific galaxy group member or gas flow.

The goal of the MUSE-ALMA Haloes project is ultimately to study the physical processes of gas transformation and flow into and out of galaxies as these are essential to a full understanding of the formation of galaxies and the growth of structure in the Universe. 

This paper outlines the overall survey strategy, describes the data acquisition and processing, and provides catalogues of all objects observed in the VLT/MUSE and HST observations of 19 quasar fields making up the MUSE-ALMA Haloes survey. The manuscript is organised as follows: Section \ref{sec:goals} presents the scientific goals of the MUSE-ALMA Haloes survey. Section \ref{sec:design} details the project's design including different observational campaigns undertaken with VLT/MUSE, ALMA and HST. Section \ref{sec:obs} focuses on the set-up of the various observing runs, while Section \ref{sec:redu} summarises the processing of these multi-wavelength datasets. In Section \ref{sec:analysis}, we provide the analysis of this dataset. The physical properties of the targets are given in Section \ref{sec:cat}. We summarize and conclude in Section \ref{sec:ccl}. We adopt an H$_0$ = 68 km s$^{-1}$ Mpc$^{-1}$, $\Omega_M$ = 0.3, and $\Omega_{\Lambda}$ = 0.7 cosmology throughout.



\section{Science Goals}
\label{sec:goals}

The MUSE-ALMA Haloes survey probes the multi-phase CGM gas of intermediate redshift galaxies ($0.19<z<1.40$). The main goal of the survey is to reveal and understand the physical
processes responsible for the rapid transformation of baryons in and out of
galaxies to address the following important questions:

\begin{itemize}

\item What is the physical relation between galaxies and their gaseous haloes? (Section \ref{sec:id})

\item How is the CGM enriched with metals? (Section \ref{sec:metallicity})

\item What is the dynamical structure of gas flows in the CGM? (Section \ref{sec:kine})

\item What is the CGM census of condensed baryons? (Section \ref{sec:census})

\end{itemize}

{We provide more details below on the global science goals of the project, which will be published in subsequent papers.  }
 


\subsection{Identify the Galaxies Associated with the Gas Traced by Absorption}
\label{sec:id}


One of the key unknowns in the study of galaxy evolution is
how galaxies acquire their gas and how they exchange this
gas with their surroundings. Historically, strong quasar absorbers were thought to be associated with isolated galaxies, likely related to their rotating disk \citep[e.g.][]{wolfe1986,wolfe2005}. While this view still partially holds, recent IFS-based findings indicate a scenario where the H\,I gas probed in absorption is related
to galaxy overdensities tracing small groups and filamentary structures \citep{peroux2019, hamanowicz2020, Dutta20, Ranchod21}. Indeed, a large fraction of the material is likely the signature of gas from remnant tidal debris
from previous interactions between the main galaxy and
other smaller satellite galaxies \citep{anglesalcazar17}. MUSE-ALMA Haloes couples absorption and emission information at high-spatial resolution over a wide-field to specifically map the relation between galaxy physical properties and extended low-surface brightness multiphase gas. Specifically, we will examine the:



i) {\it Physical properties of galaxies associated with gaseous haloes} - thanks to its wavelength coverage and high sensitivity to emission lines at optical wavelengths, the MUSE observations provide immediate confirmation of the identification of galaxies at the redshift of the absorber down to typical luminosity $L/L_{\star} \sim 0.01$. Our high-spatial resolution HST imaging complements the ground-based data by probing objects at small angular separation from the bright quasar. ALMA observations reveal molecular gas-rich objects at the redshift of the absorbers. We stress that the combination of ALMA with VLT/MUSE observations is powerful to securely assess the redshift of single-line mm detections \citep{peroux2019}. 
Together, these observations provide measurements of the impact parameter, redshift, SFR, metallicity, size, dust content, AGN contribution, stellar and molecular mass, and orientation of a sample of strong \nhi\ absorbers with log [N(H\,I)/cm$^{-2}$]$>$18. 


ii) {\it Morphology} - thanks to the high spatial resolution of the HST images, the MUSE-ALMA Haloes survey enables the study of resolved properties of the absorber host galaxy and separate interacting objects. Perturbed morphologies in galaxy groups are a signature of recent strong gravitational interactions \citep[mergers or tidal streams][]{kacprzak2007}. Additionally, the high spatial resolution afforded by the space observations resolves individual star forming clumps in the UVIS filters (resolution $\sim$0.04"). 

iii) {\it Environment} - the 1$\times$1'-field of MUSE makes it possible to establish further afield which of the galaxies are associated with the absorber \citep{Narayanan21}. Information about the environments of the absorber host galaxies distinguishes virialised groups from aligned filamentary structures and determines their typical physical scales. 




\subsection{Map the CGM Metal Distribution}
\label{sec:metallicity}


A powerful diagnostic to disentangle accreting gas from outflowing gas
around galaxies is the metallicity of the gas. Galaxy formation simulations predict infalling gas feeding galaxies from the filaments of the cosmic web to be metal-poor. Conversely, the outflowing gas is likely to be metal-enriched by the stars and supernovae within galaxies. 
While major cosmological simulations converge to predict that metallicity is strongly correlated with the direction of gas flows \citep{muratov17, peeples2019}, these projections remain essentially unconstrained observationally \citep{Wendt21}. In particular, the Illustris TNG50 and EAGLE simulations predict a strong correlation between metallicity and azimuthal angle, defined as the galiocentric angle of the quasar sightline with respect to the major axis of the central galaxy \citep{Peroux20b, VandeVoort21}. The MUSE-ALMA Haloes' survey strategy specifically combines robust gas abundance measurements (owing to known atomic gas column density N(H\,I)) with galaxy kinematics and orientation determination (thanks to the increased sensitivity at low-redshift) to investigate the following questions:

  i) {\it Gas metallicity distribution with azimuthal angle} - strong absorbers have estimates of neutral gas metallicity, [X/H], derived from multiple absorption lines in the quasar spectra (including H\,I, FeII, SiII, SII, ZnII, CrII, CIV, SiIV). The combination of VLT/MUSE and HST observations additionally provide 
 measures of the impact parameter and orientation of galaxies with respect to the absorbing gas. The azimuthal angle between the quasar line of sight and the projected galaxy's major axis on the sky is measured. 
The MUSE-ALMA Haloes analysis also puts new constraints on the metal loading factor in winds, an input to hydrodynamical simulations which limits the
amount of metals ejected by outflows \citep{nelson15a}. 

  
   ii) {\it CGM-ISM gas metallicity difference} -  the selection of absorbers with $z_{\rm abs}<$0.85 ensures that the VLT/MUSE observations deliver robust estimates of ISM metallicities measured in emission. The dataset covers
    the nebular emission lines of [O\,II] $\lambda$$\lambda$ 3727, 3730,  H$\delta$ $\lambda$ 4103, H$\gamma$ $\lambda$4342, H$\beta$ $\lambda$4863, [O\,III] $\lambda$$\lambda$4933, 5008, [N\,II] $\lambda$$\lambda$6550, 6585, H$\alpha$ $\lambda$6565 and [S\,II] $\lambda$$\lambda$ 6718, 6733 \AA\ 
    to probe the metallicity of the emitting gas based on the N2, O3N2 and/or R$_{\rm 23}$ indices as well as dust extinction. The metallicity
difference between the the absorbing gas and galaxy is positive (zero) when indicating outflow (infall).
This metallicity difference thus directly disentangles accreting metal-poor gas from outflowing metal-rich
gas \citep{peroux2016, kacprzak2019a, pointon2019}.
 
  iii) {\it Spatially resolved metallicity} - advancing from 1-dimensional metallicity gradients, MUSE-ALMA Haloes provides 2D metallicity maps to analyse the spatial distribution of metals within the ISM and CGM of galaxies with different physical properties \citep{peroux2011, rahmani2018a}. The clumpy distribution of metals is proven to be fundamental in understanding the role of interaction, mergers, accretion and gas flows in galaxy formation \citep{cresci2010, rahmani2018b, Nelson20}.

\subsection{Constrain the Dynamical Structure of Gas Flows in the CGM}
\label{sec:kine}

Determining the interactions between gas inflows 
and outflows
is important. While observational evidence for outflows is growing, direct probes of infall are notoriously difficult to gather likely because the accretion signal is swamped by that of outflows in studies of absorption back-illuminated
by the galaxy: the so-called "down-the-barrel" technique \citep{rubin2014, kacprzak2014, RobertsBorsani19, Roy21}. This method involves probing the gas lying
in front of the galaxy, with the continuum arising from the background stellar light of the galaxy. Redshifted absorption (relative to the systemic redshift) indicates inflows, i.e., the motion of
the gas towards the galaxy (or away from the observer along
the line-of-sight). Similarly, a blueshifted component suggests
outflows. While these techniques provide information on the net results of inflows and outflows, studying gas flows into and out of galaxies separately is essential for characterising their mutual interactions. Intervening quasar absorbers are uniquely suited to probe rare cases of accretion {\citep{bouche2013, rahmani2018b, ho20, szakacz21},} characterise outflows and reach low-density gas undetected by other techniques. Of particular importance is the fate of outflowing gas: escaping the galaxy potential well or recycling back to the disks via galaxy-scale fountains \citep{fraternali2008, Fraternali17, Bish19}. Studies of the multi-phase CGM are also required to better understand the physics of gas flows. With the MUSE-ALMA Haloes datasets, we will:



i) {\it Characterise accretion properties} - VLT/MUSE observations provide  information on the orientation, geometry and
kinematics of these galaxies and allow one to look for signatures of gas kinematics departing from disk rotation. Specifically, gas velocity and impact parameter measurements enable estimates of the mass flux of the accreting gas, $\dot{M}_{\rm in}$ \citep{bouche2016, zabl2019}. 
We note that measurements of the column density of the atomic gas, \nhi, are a key
ingredient to the estimates of the mass inflow rates. 


ii) {\it The fate of outflows} - identified cases of galactic winds enable robust estimates of the gas mass moving out of galaxies \citep{schroetter2019}. Early results from both VLT/SINFONI and VLT/MUSE observations indicate 
that the mass outflow rate ($\dot{M}_{\rm out}$) is similar to the star formation rate.
The outflow speeds ($\sim$100 km/s) are
smaller than the local escape velocity, which implies that the outflows do not escape the galaxy halo and are
likely to fall back into the ISM, in so-called galactic fountains \citep{schroetter2016}. On the contrary, at $z$=2--3, \cite{steidel2010} have reported that outflow speeds exceed the escape velocity at the virial radius. MUSE-ALMA Haloes will provide a large sample of new measurements of $\dot{M}_{\rm out}$. 
Such measurements
are key given that the mass loading factor, $\dot{M}_{\rm out}/SFR$ which characterises the amount of material involved in a galactic outflow and is an essential input to any theoretical model of galaxy formation - hydrodynamics, 
numerical, as well as semi-analytical models of galaxy formation \citep{Keres05, nelson15a}. 

iii) {\it Multiphase gas kinematic coupling} - high-resolution spectroscopy of the background quasars, VLT/MUSE, and ALMA observations together offer an unprecedented view of the kinematics of the atomic, ionised and molecular gas.
Our early findings \citep{Klitsch18, peroux2019, Szakacs21} indicate that while the stellar component is spatially more extended than the cold gas, the resolved molecular lines are broader in velocity space that ionised species. MUSE-ALMA Haloes provides a larger sample to further explore these results.

\subsection{Establish a Census of the Condensed CGM Baryons}
\label{sec:census}


Baryons are missing from galaxies in what is known as the {\it galaxy halo missing baryon problem} \citep{mcgaugh2008, werk2014}. These haloes lack 60~per~cent of the baryons expected from the cosmological 
mass density, suggesting missing structures both in mass and spatial extent. At $z_{\rm gal}\sim0.25$, the COS-Halos surveys indicate that the cool phase of the CGM gas accounts for half of the baryons purported to be missing from galaxy dark matter haloes \citep{ bordoloi2011, werk2013}. At $z_{\rm gal}\sim2$, the Keck Baryonic Structure Survey (KBSS) project has demonstrated from gas kinematics that 70~per~cent of galaxies with detected metal absorption have some unbound metal-enriched gas which could be a major reservoir of baryons \citep{steidel2010, rudie2019, Chen21}. There is also a SFR gap between COS-Halos and KBSS. The SFRs for COS-Halos range from $<$1M$_{\odot}$/yr to maximum of a few; the KBSS LBGs is $>$10 M$_{\odot}$/yr. 
These observed high-level of metal enrichment however have proven challenging to reproduce in simulations, solutions often relying on invoking extreme quasar-feeback mechanisms \citep[\cite{schaye15, nelson19b}, although see also][]{hafen2019}. Our MUSE-ALMA Haloes dataset is optimised to fill the "redshift gap" between COS-Halos at $z_{\rm gal}\sim0.25$ and the KBSS at $z_{\rm gal}\sim2$, and thereby to probe the effects of cosmic evolution on the metal enrichment of the CGM at the peak epoch of star formation. The MUSE-ALMA Haloes survey probes objects with SFR of a few, as the COS-Halos survey. The narrow emission lines of [O\,II],  H$\delta$, H$\gamma$, H$\beta$, [O\,III], [N\,II], H$\alpha$ and [S\,II] will avoid the systemic redshift uncertainties caused by asymmetric Ly$\alpha$ profiles seen in surveys at $z_{\rm gal}>2$. Specifically, with the MUSE-ALMA Haloes datasets, we will determine the:

i) {\it Cold gas covering fraction} - MUSE-ALMA Haloes includes measurements of the strength of the gas-phase metal (equivalent width) as a function of impact parameter and velocity separation. The optical quasar spectroscopy delivers accurate equivalent width measurements of Mg\,II to probe the metal distribution of CGM about galaxies as a function of impact parameter \citep{Dutta20, Dutta21}. The larger number of objects in the sample enables us to quantify these results as a function of galaxy properties (redshift, SFR, metallicity, mass, environment). 
Indeed, the remarkable combination of VLT/MUSE's unmatched sensitivity and wide FoV makes the instrument an efficient "redshift machine" to build a sizeable sample of Mg\,II and [O\,II] emitters at $z\leq0.8$. 

ii) {\it Baryonic and gas fractions of galaxies} - most recent CO-emission surveys have informed us about the gas fraction of galaxies up to $z\sim$4 \citep[COLD GASS and PHIBSS, respectively][]{saintonge2017, tacconi2018, Tacconi20}. Yet these results are limited to only the most massive galaxies. MUSE-ALMA Haloes characterises the gas fraction of the bulk of the galaxy population by reaching objects with stellar masses M$_*<$10$^{10}$ M$_{\odot}$. These gas mass measurements provide the determination of both the baryonic fraction $f_{\rm baryons}=(M_{\rm gas}+M_{*})/M_{\rm dyn}$, and gas fraction $\mu=M_{\rm gas}/(M_{\rm gas}+M_{*})$ in a population of galaxies significantly fainter than probed otherwise.

iii) {\it Galactic baryon cycling} - MUSE-ALMA Haloes probes the molecular masses from the flux of CO emission detected with ALMA. Combined with known SFR estimates, we calculate the molecular depletion times as: $\tau_{\rm depl}$=$M_{\rm H2}$/SFR=1/SFE, where SFE is the star formation efficiency, a key link to the baryon cycle \citep{Peroux20b, Walter20}.
\section{Survey Design}
\label{sec:design}

The MUSE-ALMA Haloes survey firstly aims to study the galaxies associated with strong gas absorbers. With this goal in mind, a set of 19 quasar fields were selected to have so-called "primary targets". On top of this unique sample, numerous additional galaxies of interest are also included forming the so-called "galaxy-selected" sample.

\begin{table*}
\begin{center}
\caption{{\bf Primary targets within the 19 fields targeted by the MUSE-ALMA Haloes survey.} The sky coordinates, redshift and V-band magnitude of the background quasars are provided. The table also lists the redshifts and neutral gas column densities, \logNHI, as estimated from HST UV-spectra of the main targets of the survey, the so-called primary sample. {The penultimate column lists the HST instrument used (FOS, COS or STIS).} Indications of additional ground-based high-resolution spectroscopy of the background quasar are also provided if available in the last column (Keck/HIRES, VLT/UVES or VLT/X-Shooter).
}
\begin{tabular}{llcccccccc}
\hline\hline
Quasar 		 &Other Name &RA &Dec  &$z_{\rm quasar}$  & V mag &$z_{\rm abs}$ &$\logNHI$ &UV  &Optical \\
                 & &(J2000)   &(J2000)     &                &   & &   & QSO spe & QSO spec\\
\hline                                                        
Q0058$+$0019 &LBQS0058$+$0155  &01 00 54.12 &$+$02 11 36.33 &1.96  &17.16  &0.6125 &{20.08$\pm$0.15$^a$}  &{FOS}   &UVES \\
...  &...    &... &... &...  &... &...                &...                                                                 &{...}   &$+$HIRES\\
Q0123$-$0058 &...               &01 23 03.22 &$-$00 58 19.38 &1.55  &18.75 &0.8686 &{$<$18.62$^b$}        &{STIS}  &UVES  \\
...  &...    &... &... &...  &...                                          &1.4094 &{20.08$\pm$0.09$^c$}  &{...}   &...\\
Q0138$-$0005 &...               &01 38 25.49 &$-$00 05 33.97 &1.34  &18.76 &0.7821 &{19.81$\pm$0.08$^d$}  &{STIS}  &UVES  \\
J0152$-$2001 &UM 675           &01 52 27.34 &$-$20 01 07.10 &2.06  &17.4   &0.3830&{$<$18.78$^e$}        &{FOS}   &HIRES \\
...  &...    &... &... &...  &...                                          &0.7802 &{18.87$\pm$0.12$^b$}  &{...}   &...\\
Q0152$+$0023 &...               &01 52 49.68 &$+$00 23 14.60 &0.59  &17.87 &0.4818 &{19.78$\pm$0.08$^b$}  &{STIS}  &...    \\
Q0420$-$0127 &J0423$-$0130     &04 23 15.80 &$-$01 20 33.07 &0.91  &17.00  &0.6331 &{18.54$\pm$0.09$^b$}  &{FOS}   &HIRES \\ 
Q0454$+$039  &Q0454$+$0356     &04 56 47.17 &$+$04 00 52.94 &1.34  &16.53  &0.8596 &{20.67$\pm$0.03$^b$}  &{STIS}  &UVES \\
...  &...    &... &... &...  &...                                          &1.1532 &{18.59$\pm$0.02$^b$}  &{...}   &$+$HIRES\\
Q0454$-$220  &J0456$-$2159     &04 56 08.92 &$-$21 59 09.60 &0.53  &16.10  &0.4744 &{19.45$\pm$0.03$^b$}  &{COS}   &UVES\\
...  &...    &... &... &...  &...                                          &0.4833 &{18.65$\pm$0.02$^b$}  &{STIS}  &$+$HIRES\\
Q1110$+$0048 &2QZJ1110$+$0048  &11 10 55.00 &$+$00 48 54.40 &0.76  &18.68  &0.5604 &{20.20$\pm$0.10$^b$}  &{STIS}  &...    \\ 
J1130$-$1449 &B1127$-$145      &11 30 07.04 &$-$14 49 27.40 &1.19  &16.90  &0.1906 &{$<$19.10$^f$}        &{FOS}   &UVES  \\
...  &...    &... &... &...  &...                                          &0.3127 &{21.71$\pm$0.08$^b$}  &{...}   &...\\
...  &...    &... &... &...  &...                                          &0.3283 &{$<$18.90$^f$}       &{...}   &...\\
J1211$+$1030 &LBQS1209$+$1046  &12 11 40.59 &$+$10 30 02.00 &2.19  &18.37  &0.3929 &{19.46$\pm$0.08$^b$}  &{FOS}   &UVES  \\
...  &...    &... &... &...  &...                                          &0.6296 &{20.30$\pm$0.24$^b$}  &{...}   &...\\
...  &...    &... &... &...  &...                                          &0.8999 &{$<$18.50$^f$}        &{...}   &...\\
...  &...    &... &... &...  &...                                          &1.0496 &{$<$18.90$^f$}        &{...}   &...\\
Q1229$-$021  &Q1232$-$0224     &12 32 00.01 &$-$02 24 04.80 &1.05  &17.06  &0.3950 &{20.75$\pm$0.07$^g$}  &{COS}   &UVES  \\
...  &...    &... &... &...  &...                                          &0.7572 &{18.36$\pm$0.09$^b$}  &{...}   &...\\
...  &...    &... &... &...  &...                                          &0.7691 &{18.11$\pm$0.15$^f$}  &{...}   &...\\
...  &...    &... &... &...  &...                                          &0.8311 &{18.84$\pm$0.10$^f$}  &{...}   &...\\
Q1342$-$0035 &LBQS1340$-$0020  &13 42 46.23 &$-$00 35 44.27 &0.79  &18.30  &0.5380 &{19.78$\pm$0.13$^b$}  &{STIS}  &...    \\
Q1345$-$0023 &LBQS1343$-$0008  &13 45 47.82 &$-$00 23 23.86 &1.10  &17.60  &0.6057 &{18.85$\pm$0.20$^b$}  &{STIS}  &...    \\
Q1431$-$0050 &LBQS1429$-$0036  &14 31 43.74 &$-$00 50 12.48 &1.18  &18.23  &0.6085 &{19.18$\pm$0.24$^b$}  &{STIS}  &...    \\
...  &...    &... &... &...  &...                                          &0.6868 &{18.40$\pm$0.07$^b$}  &{...}   &...\\
Q1515$+$0410 &...               &15 15 05.12 &$+$04 10 12.10 &1.27  &18.59 &0.5592 &{20.20$\pm$0.19$^h$}  &{STIS}  &XSH    \\
Q1554$-$203  &J1557$-$2029     &15 57 21.18 &$-$20 29 12.10 &1.95  &19.2   &0.7869 &{$<$19.00$^b$}        &{STIS}  &XSH    \\
J2131$-$1207 &B2128$-$123      &21 31 35.26 &$-$12 07 04.80 &0.50  &16.11  &0.4298 &{19.50$\pm$0.15$^i$}  &{STIS}  &UVES \\
...  &...    &... &... &...  &... &...                &...                                                                 &{...}   &$+$HIRES\\
Q2353$-$0028 &LBQS2350$-$0045  &23 53 21.61 &$-$00 28 41.66 &0.76  &18.13  &0.6044 &{21.54$\pm$0.15$^b$}  &{STIS}  &...    \\
\hline\hline 				       			 	 
\label{tab:abs}
\end{tabular}			       			 	 
\begin{minipage}{180mm}
Note: HIRES = Keck/HIRES; UVES = VLT/UVES; XSH = VLT/X-Shooter\\
  {References:
  $^a$ \cite{pettini2000}
  $^b$ \cite{rao2006}
  $^c$ \cite{meiring2009}
  $^d$ \cite{peroux2008}
  $^e$ \cite{rahmani2018a}
  $^f$ \cite{hamanowicz2020}
  $^g$ \cite{boisse1998}
  $^h$ \cite{rahmani2016}
  $^i$ \cite{muzahid2016}
  }
\end{minipage}
\end{center}			       			 	 
\end{table*}

\begin{figure*}
	\includegraphics[width=8.5cm]{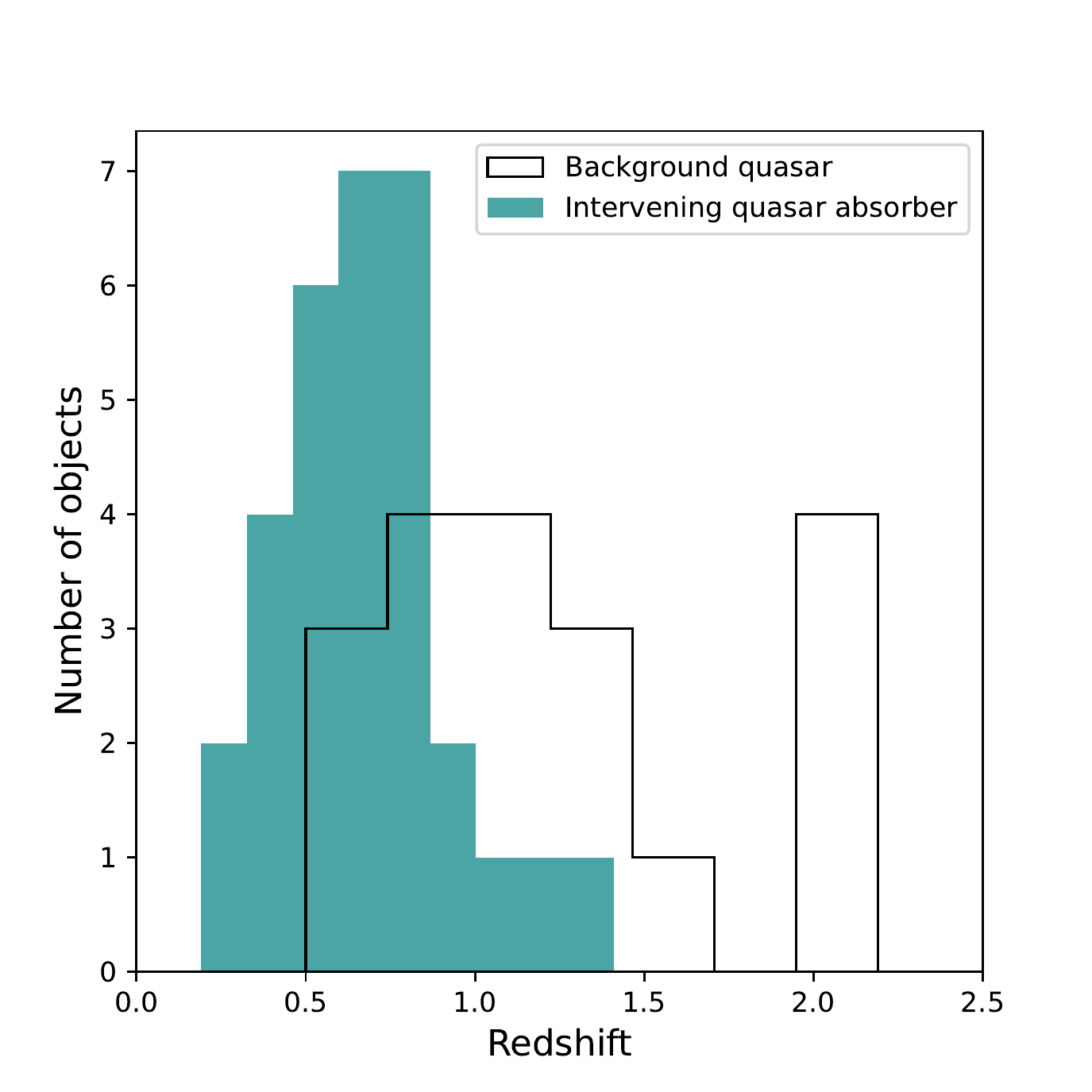}
	\includegraphics[width=8.5cm]{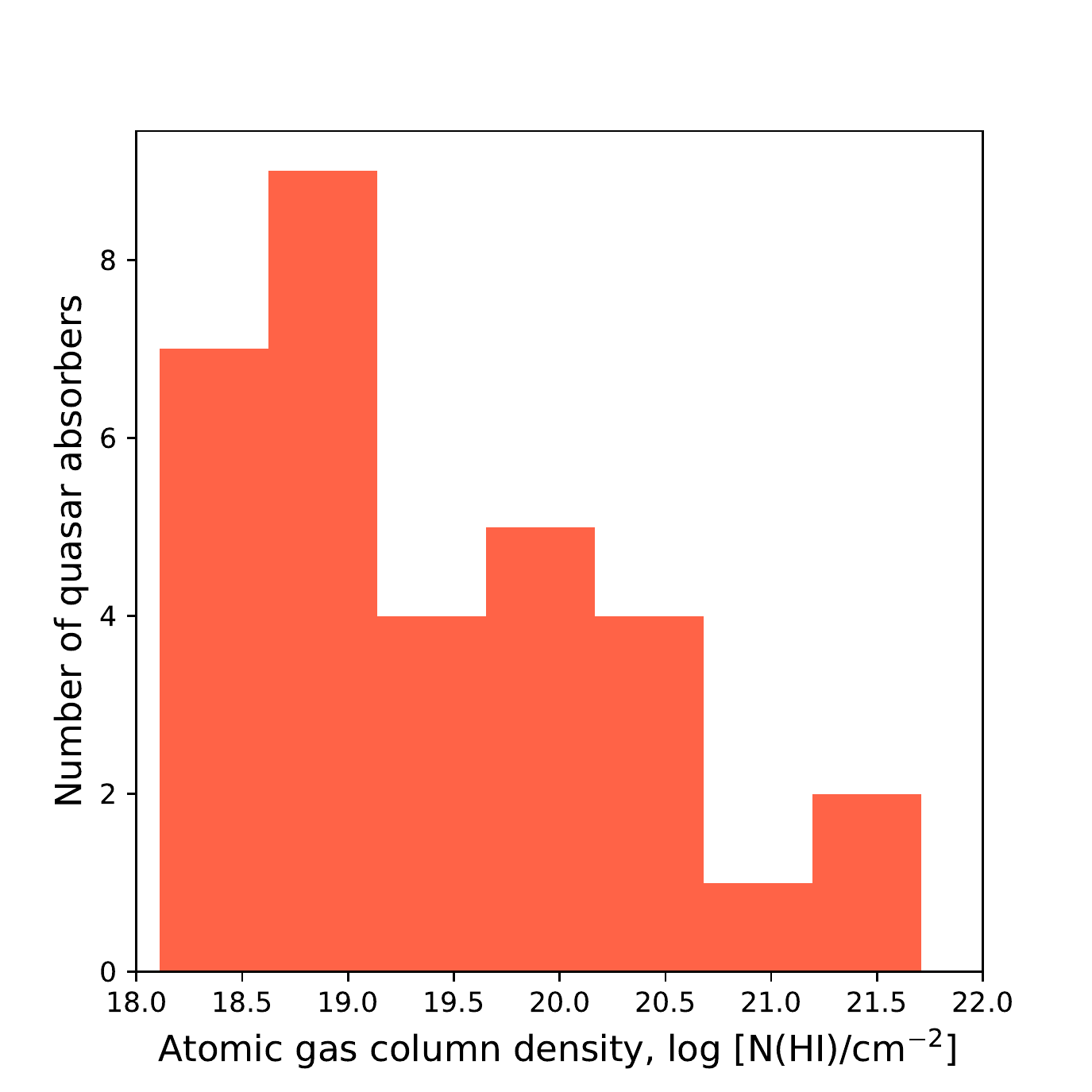}
    \caption{{\bf Quasars and primary targets properties.} {\it Left panel:} Redshift distribution (in black) of the 19 background quasars making up the MUSE-ALMA Haloes survey. The filled histogram (green) displays the redshift distribution of the 32 intervening absorbers along the line-of-sight to these quasars, composing the primary target sample. {\it Right panel:} Atomic hydrogen gas column density distribution of these strong absorbers. The selection criterion imposes the cut at log [N(H\,I)/cm$^{-2}$]$>$18. }
    \label{fig:zabs_histo}
\end{figure*}

\subsection{The Primary Targets}

MUSE-ALMA Haloes is based on a unique selection of known quasar absorbers based solely on two criteria:

\begin{enumerate}
    \item measured \hi\ column density log [N(H\,I)/cm$^{-2}$]$>$18, from HST UV spectroscopy, which delivers spectra with resolutions R=20,000-30,000 
    
    \item $z_{\rm abs} <0.85$, to ensure that all emission lines up to [OIII] $\lambda$ 5007\AA\ will be covered by the VLT/MUSE observations
    
\end{enumerate}

MUSE-ALMA Haloes complements other surveys thanks to these key elements of the selection. The \hi\ upper limits denote quasar absorbers with columns very near this limit, and in all cases, well above log [N(H\,I)/cm$^{-2}$]$>$15. {The observations were undertook with either HST/FOS \citep{Harms91}, COS \citep{Green12} or STIS \citep{Kimble98}. References to the published \hi\ column density measurements are provided as a footnote to Table~\ref{tab:abs}.} Indeed, we stress that an accurate knowledge of \hi\ column density in absorption is pivotal to a precise measure of neutral gas metallicity including a potentially required correction for the photoionised fraction of the gas. 
In addition, \hi\ is a key ingredient in the estimate of the mass loading factor. The redshift range, which complements other efforts at $z>3$, is set to permit the robust determination of the systemic redshift based on narrow emission lines (as opposed to Ly$\alpha$), as well as the star formation rates and emission metallicity of these galaxies based on rest-frame optical emission diagnostics. 
Finally, the larger apparent size and surface brightness at these relatively modest redshifts compare to higher-redshifts enable a kinematic reconstruction of a large number of objects. The sample resulting from this selection comprises 32 individual absorbers (the "primary targets") in 19 unique fields. The \nhi\ and redshifts of these systems are provided in Table~\ref{tab:abs}. The distributions of their redshifts and \hi\ columns are presented in Fig.~\ref{fig:zabs_histo}.

The selected primary targets are well-studied strong absorbers. For this reason, they benefit from a number of addition ancillary data sets. By selection, all fields have high-resolution UV spectroscopy from HST. Most of the targets have ground-based optical high-resolution quasar spectra from VLT/UVES, X-Shooter or Keck/HIRES observations. These spectra provide abundance estimates based on the weaker metal lines of e.g. Si\,II, S\,II, Zn\,II and Cr\,II. One of the absorbers (towards Q1229$-$021) has also been detected in 21cm against the radio-loud background quasar as part of the FLASH survey currently on-going on ASKAP \citep{Sadler20}. More of the MUSE-ALMA Haloes targets will also be part of FLASH's Phase 2, potentially providing additional information on atomic gas kinematics and spin temperature. While this paper focuses on the newly acquired data, the ancillary observations will be presented in upcoming publications. The instruments used to record the high-resolution quasar spectra are also provided in Table \ref{tab:abs}. 
 
 \subsection{Additional Absorbers}

In addition to the primary targets, with log [N(H\,I)/cm$^{-2}$]$>$18, the sightlines to the background quasars contain a number of additional metal absorbers and higher-redshift systems. While these are not directly the targets of the primary sample, most of the science described in Section~\ref{sec:goals} is also addressed by these additional absorbers. These typically include MgII absorbers with log [N(H\,I)/cm$^{-2}$]$<$18 and associated with well-identified [OII] emitters observed with VLT/MUSE \citep[see e.g.][]{rahmani2018b,hamanowicz2020}. We refer to this sample as the "additional absorbers".

\subsection{The Galaxy-selected Sample}

Beyond the absorber-selected targets, each of the 19 VLT/MUSE cubes contains several dozen $z_{\rm gal}<1.2$ galaxies with secure redshifts below the quasar redshift (typically $<z_{\rm quasar}>=1.2$). This additional sample includes 215 galaxies across all fields. These objects are selected independently of whether they are related to a known absorber, but they are at redshifts for which we have coverage in the existing quasar spectra to characterise the cold gas traced by e.g. Mg\,II absorbers. This provides a large sample of {\it absorber-blind} galaxies with a wealth of information about their physical properties. Indeed, statistically probing the physical properties of galaxies which do not have extended gaseous haloes is important to draw a full picture of the baryon cycle \citep{Chen21}. Equally, the HST observations provide a measure of the stellar mass of several hundreds of galaxies, including numerous objects outside the VLT/MUSE field-of-view. We refer to this part of the survey as the "galaxy-selected sample". This larger dataset constitutes the main focus of the current paper. We note that by construction the primary sample is a subset of the galaxy-selected sample. 

\subsection{Supplementary Science Targets}

In addition to the main science cases, the MUSE-ALMA Haloes data provide detailed information for 19 bright $0.48<z_{\rm quasar}<2.19$ quasars. The dataset includes high spatial resolution ($\sim$2kpc) observations of potential quasar-driven massive outflows thought to be omnipresent in bright quasar-host galaxies \citep{harrison16, harrison18}. Moreover, MUSE-ALMA Haloes covers a large volume robust against cosmic variance, providing spectra and redshifts for hundreds of galaxies ranging from $z=0$ to $z=6.5$, including Ly$\alpha$ emitters at $z_{\rm gal}>3$.


\section{Observing Strategy}
\label{sec:obs}

\subsection{Optical Integral Field Spectrograph VLT/MUSE Observations}

The VLT/MUSE observations were carried out in service mode (under programmes ESO 96.A-0303, 100-A-0753, 101.A-0660 and 102.A-0370, PI: C. P\'eroux and 298.A-0517, PI: A. Klitsch) at the European Southern Observatory on the 8.2 m Yepun telescope. Each MUSE observations were centered on the bright background quasar. To optimise the schedulability of the runs, a mix of natural seeing mode and GALACSI Adaptive Optics (AO) system observations was used as indicated in Table \ref{tab:JoO_MUSE}. In addition, to ease the scheduling of the P100, P101 and P102 runs, we adopted a proven strategy tested during our early study of relaxing the observing conditions where the gain from AO is the highest. The resulting image quality (FWHM values) are listed in Table \ref{tab:JoO_MUSE}. The table also includes exposure times for each field observed. Each exposure was further divided into two equal sub-exposures, with an additional field rotation of 90 degrees and sub-arcsec dithering offset in 2-step pattern to minimise residuals from the slice pattern. The resulting field of view is 59.9 arcsec $\times$ 60 arcsec, with a 0.2 arcsec/pixel scale. We used the "nominal mode" resulting in a spectral coverage of $\sim$4800-9300 \AA\ {\citep{bacon2010}}. The AO-assisted observations are blind to wavelengths cut-out by the notch filter between 5820-5970 \AA. The spectral resolution is R=1770 at 4800 \AA\ and R=3590 at 9300 \AA\ resampled to a spectral sampling of 1.25 \AA/pixel. A journal of observations summarising the properties of the MUSE program is presented in Table~\ref{tab:JoO_MUSE}. 

\begin{table*}
\begin{center}
\caption{{\bf Journal of VLT/MUSE observations.} The MUSE observations were carried out in service mode over various observing campaigns with a combination of natural seeing and GALACSI Adaptive Optics (AO) system modes. 
{
{The $3\sigma$ line flux limits are calculated at $7000$\ \AA \ for an unresolved source spread over a disk with diameter equal to the seeing and with FWHM $= 3$\ \AA.}}
}
\begin{tabular}{lcclcccccc}
\hline\hline
Quasar 		 &$t_{\rm exp}$ &AO? &I.Q.$^a$ & {Flux limit} &Prog. ID &PI &MUSE-ALMA Haloes\\
field                 &[s]    &         &["] & [$\text{erg s}^{-1} \text{cm}^{-2}$] & &&references\\
\hline                                                        
Q0058$+$0019 &1410+330                      &AO &1.23  & 2.8$\times 10^{-17}$  &102.A-0370  &P\'eroux &this work\\
Q0123$-$0058 &1410$\times$2                 &no &2.11  & 5.5$\times 10^{-17}$ &100.A-0753  &P\'eroux &this work\\
Q0138$-$0005 &1410$\times$2                 &AO &1.11  & 3.6$\times 10^{-17}$ &101.A-0660  &P\'eroux &this work\\
J0152$-$2001 &1200$\times$4                 &no &0.72  & 4.7$\times 10^{-18}$ &096.A-0303  &P\'eroux &\cite{rahmani2018a,rahmani2018b}\\
Q0152$+$0023 &1410$\times$2                 &AO &0.65  & 1.2$\times 10^{-17}$  &101.A-0660  &P\'eroux &this work\\
Q0420$-$0127 &1405$\times$4                 &no &0.71  & 4.0$\times 10^{-18}$ &298.A-5017  &Klitsch  &\cite{Klitsch18}\\
Q0454$+$039  &1410$\times$4                 &no &0.81  & 5.3$\times 10^{-18}$ &100.A-0753  &P\'eroux &this work\\
Q0454$-$220  &1410$\times$2                 &no &0.57  & 2.3$\times 10^{-18}$ &100.A-0753  &P\'eroux &this work\\
Q1110$+$0048 &1410$\times$2                 &AO &0.52  & 2.6$\times 10^{-18}$ &101.A-0660  &P\'eroux &this work\\
J1130$-$1449 &1200$\times$6+960$\times$2    &no &0.76  & 4.2$\times 10^{-18}$ &096.A-0303  &P\'eroux &\cite{peroux2019}\\
J1211$+$1030 &1200$\times$4                 &no &0.75  & 6.8$\times 10^{-18}$ &096.A-0303  &P\'eroux &\cite{hamanowicz2020}\\
Q1229$-$021  &1200$\times$4                 &no &0.80  & 7.3$\times 10^{-18}$ &096.A-0303  &P\'eroux &\cite{hamanowicz2020}\\
Q1342$-$0035 &1410$\times$2                 &AO &1.30  & 2.0$\times 10^{-17}$ &101.A-0660  &P\'eroux &this work\\
Q1345$-$0023 &1410$\times$2                 &AO &0.64  & 4.8$\times 10^{-18}$ &101.A-0660  &P\'eroux &this work\\
Q1431$-$0050 &1410$\times$2                 &AO &0.54  & 7.1$\times 10^{-18}$ &101.A-0660  &P\'eroux &this work\\
Q1515$+$0410 &1410$\times$4                 &AO &0.59  & 5.2$\times 10^{-18}$ &101.A-0660  &P\'eroux &this work\\
Q1554$-$203  &1410$\times$4                 &AO &0.79  & 6.0$\times 10^{-18}$ &101.A-0660  &P\'eroux &this work\\
J2131$-$1207 &1200$\times$4                 &no &0.72  & 4.6$\times 10^{-18}$ &096.A-0303  &P\'eroux &\cite{peroux2017, Szakacs21}\\
Q2353$-$0028 &1410$\times$4                 &AO &0.81  & 5.2$\times 10^{-18}$ &101.A-0660  &P\'eroux &this work\\
\hline\hline 				       			 	 
\label{tab:JoO_MUSE}
\end{tabular}			       			 	 
\begin{minipage}{180mm}
Note: $^a$ I.Q. refers to the image quality in the reconstructed cube measured from a Gaussian fit at 7000 \AA.
\end{minipage}
\end{center}			       			 	 
\end{table*}			       			 	 

\subsection{ALMA mm Observations}

The ALMA data included in the MUSE-ALMA Haloes survey are comprised of three distinct catalogues: i) proposals led by our group, ii) targets included in ALMACAL and iii) archival data. 

Firstly, a subset of the quasar fields were observed with ALMA specifically for use in the MUSE-ALMA Haloes survey. The details of these runs are described in \cite{peroux2019, Klitsch21, Szakacs21} and we only give a brief overview here. The observations were performed in Bands 3, 4 or 6 and covered the CO(1--0), CO(2--1) or CO(3--2) emission lines at the redshift of the primary targets. The programmes were 2016.1.01250.S and 2017.1.00571.S (PI: C. Péroux) and 2018.1.01575.S (PI: A. Klitsch). The precipitable water vapour (PWV) for these observations varied between 0.65 and 5.4 mm and the total on-source observing times are listed in Table~\ref{tab:JoO_ALMA}. The data were observed in relatively compact antenna configurations which resulted in an angular resolution of the order 1\arcsec. For each target, one of the four spectral windows was centred on the redshifted CO line frequency and used relatively high spectral resolution (4096 dual-polarization channels across a bandwidth of 1875~MHz) whilst the other three spectral windows were used to observe the continuum and thus only required 128 channels each over the same bandwidth.

Second, some of the MUSE fields are part of the ALMACAL\footnote{\url{almacal.wordpress.com}} survey. ALMACAL is an ingenious mm survey that exploits ALMA phase and amplitude calibration data.
Since 20~per~cent of all ALMA telescope time has been spent on calibrators, ALMACAL is already the widest and deepest mm survey. The amount of data processed to date adds up to over 2000 hrs of ALMA observation time, equivalent to about half of all observing time awarded to observers in a one-year ALMA cycle. 
These data become publicly available immediately, without proprietary time. This calibrator survey can be used to study both the calibrators themselves and any serendipitously-detected galaxies in the field. One of the fields (namely Q0420$-$0127) had already appropriate frequency coverage and sufficient depth to enable a detailed study \citep{Klitsch18}. 
As ALMA is observing repeatedly the same calibrator fields, further data will likely become available in the future. 

Thirdly, data for four of the targets are available in the ALMA archive, two of which (Q0058$+$0019 and Q0138$-$0005)  were previously published \citep{kanekar2018}. 

A journal of observations summarising the properties of the ALMA runs {as of July 15$^{\rm th}$ 2022} 
is presented in Table~\ref{tab:JoO_ALMA}. The table includes exposure times, spatial resolution, primary beam diameter and frequency coverage.

\begin{table*}
\begin{center}
  \caption{{\bf Journal of ALMA observations.}
    "PI data" are from our group, "ALMACAL" means that the target is observed as an ALMA calibrator\protect\footnote{almacal.wordpress.com/}\protect and  "archive" refers to the general ALMA archive (shown in {\it italics}) as of July 15th 2022. 
    "..." refers to entry with no ALMA observations.}
\begin{tabular}{lllllllllll}
\hline\hline
Quasar &$t_{\rm exp}$  &Catalogue &Resolution &ALMA &Primary  &Frequency &{Cont.} &Prog. ID &PI &References\\
field                 &    &        & &Band  &Beam &cov  &{Sens} &&\\
               &[min]    &        &["] &  &["] &[GHz]  &{[mJy]} &&\\
\hline                                                        
Q0058$+$0019 &41    &{\it archive}   &1.61    &4   &44    &129.91-145.89   &{0.014}   &2013.1.01178.S &Prochaska &\cite{kanekar2018}\\
...          &66    &{\it archive}   &1.76    &4   &44    &129.97-145.89   &{0.017}   &2015.1.01034.S &Prochaska &\cite{kanekar2018}\\  
Q0123$-$0058 &...   &...             &...     &... &...   &...             &{...}   &...                  &...       &...\\
Q0138$-$0005 &12    &{\it archive}   &1.63    &4   &44    &128.43-143.99   &{0.029}   &2013.1.01178.S &Prochaska &\cite{kanekar2018}\\
...          &32    &{\it archive}   &0.08    &4   &44    &128.43-143.99   &{0.013}   &2015.1.01034.S &Prochaska &\cite{kanekar2018}\\  
J0152$-$2001 &118   &PI data         &0.90    &6   &27    &232.08-251.05   &{...}   &2017.1.00571.S &P\'eroux  &\cite{Szakacs21}\\
Q0152$+$0023 &...   &...             &...     &... &...   &...             &{...}   &...                  &...       &...\\
Q0420$-$0127 &1303  &ALMACAL         &1.75    &3   &62    &84.03-115.88    &0.0035   &...                  &...       &...\\
...          &260   &ALMACAL         &0.77    &4   &44
 &125.03-162.88   &0.0071   &...                  &...       &\cite{Klitsch18}\\ 
...          &256   &ALMACAL         &1.29    &5   &33
 &163.28-207.54   &0.010   &...                  &...       &...\\
...          &2167  &ALMACAL         &0.57    &6   &27
 &211.09-274.99   &0.0050   &...                  &...       &\cite{Klitsch18}\\ 
...          &1332  &ALMACAL         &0.35    &7   &18
 &277.00-685.56   &0.0078   &...                  &...       &...\\
...          &162   &ALMACAL         &0.25    &8   &12
 &396.50-498.65   &0.0090   &...                  &...       &...\\
...          &20     &ALMACAL         &0.19    &9   &9
 &657.66-694.75   &0.5   &...                  &...       &...\\
...          &2.5   &{\it archive}   &1.76    &3   &62    &88.20-90.69     &{2.3}   &2015.1.00503.S &Bronfman  &...\\
...          &4.5   &{\it archive}   &1.13    &3   &62    &112.51-115.89   &{3.1}   &2015.1.00503.S &Bronfman  &...\\  
...          &7.5   &{\it archive}   &1.79    &3   &62    &88.25-91.13     &{1.2}   &2019.1.00743.S &Finger    &...\\  
...          &2.0   &{\it archive}   &2.22    &3   &62    &113.06-115.64   &{3.5}   &2019.1.00743.S &Finger    &...\\  
Q0454$+$039  &...   &...             &...     &... &...    &...             &{...}   &...                 &...       &...\\
Q0454$-$220  &...   &...             &...     &... &...    &...             &{...}   &...                 &...       &...\\
Q1110$+$0048 &...   &...             &...     &... &...    &...             &{...}   &...                 &...       &...\\
J1130$-$1449 &227   &PI data         &1.07    &3   &62    &86.88-102.80    &{0.0098}   &2016.1.01250.S  &P\'eroux  &\cite{peroux2019}\\
...          &57    &ALMACAL         &0.79    &3   &62    &85.91 -113.63   &{0.018}   &...                 &...       &...\\
...          &19    &ALMACAL         &1.01    &4   &44
 &131.57-151.98   &{0.029}   &...                 &...       &...\\
...          &8     &ALMACAL         &0.98    &6   &27
 &211.98-245.83   &{0.057}   &...                 &...       &...\\
...          &9     &ALMACAL         &0.29    &7   &18
 &335.50-358.93   &{0.080}   &...                 &...       &...\\
J1211$+$1030 &45    &PI data         &0.69    &6   &27    &231.99-250.97   &{0.014}   &2017.1.00571.S &P\'eroux  &\cite{Szakacs21}\\
Q1229$-$021  &293   &PI data         &0.99    &6   &27    &246.95-265.99   &{0.018}   &2017.1.00571.S &P\'eroux  &\cite{Szakacs21}\\
...          &50    &ALMACAL         &1.91    &3   &62
 &89.44 -115.08    &{0.016}            &...     &...       &...\\
...          &34    &ALMACAL         &0.41    &6   &27
 &213.99-266.70    &{0.022}            &...     &...       &...\\
...          &3     &ALMACAL         &0.14    &7   &18
 &335.50-351.48    &{0.071}            &...     &...       &...\\
...          &3     &{\it archive}   &0.67    &6   &27   &223.01-242.99     &{0.052}      &2015.1.00932.S    &Meyer      &...\\  
...          &1     &{\it archive}   &0.89   &3   &62   &89.50-105.48       &{0.059}       &2016.1.01481.S     &Meyer     &...\\  
...          &2     &{\it archive}   &0.44   &6   &27   &222.99-242.97      &{0.062}       &2016.1.01481.S     &Meyer     &...\\  
...          &4     &{\it archive}   &0.35   &3   &62   &89.50-105.48       &{0.035}       &2015.1.00932.S     &Meyer     &...\\  
...          &7     &{\it archive}   &0.20   &6   &27   &223.03-243.01      &{0.032}       &2016.1.01481.S    &Meyer     &...\\  
...          &4     &{\it archive}   &0.22   &3   &62      &89.51-105.49    &{0.030}       &2016.1.01481.S     &Meyer     &...\\  
Q1342$-$0035 &...   &...             &...     &... &...    &...             &{...}        &...             &...       &...\\
Q1345$-$0023 &...   &...             &...     &... &...    &...             &{...}        &...             &...       &...\\
Q1431$-$0050 &...   &...             &...     &... &...    &...             &{...}        &...             &...       &...\\
Q1515$+$0410 &...   &...             &...     &... &...    &...             &{...}        &...             &...       &...\\
Q1554$-$203  &...   &...             &...     &... &...    &...             &{...}        &...             &...       &...\\
J2131$-$1207 &120   &PI data         &0.97    &6   &27    &224.00-242.80    &{0.013}   &2017.1.00571.S    &P\'eroux  &\cite{Szakacs21}\\   
...          &6     &PI data         &0.65    &4   &44    &146.55-162.17    &{0.033}   &2018.1.01575.S    &Klitsch   &\cite{Klitsch21}\\  
...          &4     &ALMACAL         &1.09    &3   &62
 &88.36 -105.49    &{0.022}            &...     &...       &...\\
...          &10    &ALMACAL         &0.78    &6   &27
 &213.53-269.93    &{0.011}            &...     &...       &...\\
...          &2     &ALMACAL         &0.46    &7   &18
 &335.50-351.49    &{0.090}            &...     &...       &...\\
Q2353$-$0028 &...   &...             &...     &... &...    &...                 &{...}            &...     &...       &...\\
\hline\hline        
\label{tab:JoO_ALMA}
\end{tabular}        
\end{center}        
\end{table*}

\subsection{HST Broad-Band Imaging}

MUSE-ALMA Haloes also includes broad-band imaging
of most of the fields in the sample. The new data were observed during Cycle 27 as part of a 40-orbit medium programme (ID: 15939;
PI: C. P\'eroux) with the Wide Field Camera 3 in both the optical (UVIS) and infrared (IR) detectors, using different combination of
broad-band filters as indicated in Table \ref{tab:JoO_HST}. We also make use of archival data recorded with WFPC.
Together, the observations took place between January 2015 and May 2021. For the dedicated programme, we aimed at setting the roll-angle of the telescope such that the primary target galaxy counterpart revealed by MUSE lies at 45 degrees from the diffraction spikes of the instrument Point Spread
Function (PSF). We use a dithering pattern in four individual exposures to help with removal of cosmic rays and hot pixels. The UVIS observations were initially taken using the WFC3-UVIS-DITHER-BOX pattern. The two observations with the IR detector were taken using the WFC3-IR-DITHER-BOX-MIN pattern providing an optimal 4-point sampling of the PSF. Some of the observations failed due to HST pointing drifts and only a fraction of these fields were reobserved. In these cases, the dithering pattern might differ slightly. For some fields, we also rely on existing archival data to constrain the
galaxy spectral energy distribution (SED) on either side of the 4000 \AA\ Balmer break. The majority (16/19) of the fields have three broad-band filters available. Using additional data available in the HST archive, we even cover a total of 4 filters for some of the targets. A summary of the observational set-up, including observing times, is given in Table~\ref{tab:JoO_HST}. 

\begin{table*}
\begin{center}
\caption{{\bf Journal of HST Observations.} The broad-band imaging observations available from the Space Telescope Science Institute archive are shown in {\it italics}. In the cases of several primary targets in the field, the chosen broad-band filters are well suited to probe the wavelength range bluewards of the 4000\AA\ break.  "..." refers to entry with no HST observations. }
\begin{tabular}{lllllll}
\hline\hline
Quasar field	&t$_{\rm exp}$  	  &HST  &{Central $\lambda$} &{Mag} &Prog. ID &PI \\
		&[s]	  &camera/filter &{[\AA]} &{Limit} &&		 	 \\
\hline                                                        
Q0058$+$0019 &27676   &{\it WFPC F702W}    &6919        &27.02              &6557  &Steidel   \\ 
Q0123$-$0058 &...     &...                 &...         &...                &...   &...       \\
Q0138$-$0005 &2367    &WFC3/UVIS F475W	   &4774	    &28.70             &15939 &P\'eroux  \\
...          &2364    &WFC3/UVIS F625W	   &6242	    &28.90             &15939 &P\'eroux  \\
...          &2412    &WFC3/IR F105W	   &10557	    &27.93             &15939 &P\'eroux  \\
J0152$-$2001 &2364    &WFC3/UVIS F336W	   &3355	    &29.39             &15939 &P\'eroux  \\               
...          &2364    &WFC3/UVIS F475W	   &4774	    &29.29             &15939 &P\'eroux  \\
...          &1250    &{\it WFPC F702W}    &6919        &27.68                     &6557  &Steidel   \\
Q0152$+$0023 &2106    &WFC3/UVIS F336W	   &3355	    &28.54             &15939 &P\'eroux  \\
...          &2489    &WFC3/UVIS F475W	   &4774	    &28.39             &15939 &P\'eroux  \\
...          &2489    &WFC3/UVIS F814W	   &8033	    &27.92             &15939 &P\'eroux  \\
Q0420$-$0127 &2122    &WFC3/UVIS F336W	   &3355	    &28.30             &15939 &P\'eroux  \\
...          &2216    &WFC3/UVIS F475W	   &4774	    &29.52             &15939 &P\'eroux  \\
...          &1918    &{\it NICMOS F160W}  &1607        &24.12              &7451  &Smette    \\
Q0454$+$039  &2416    &WFC3/UVIS F625W	   &6242	    &28.72             &15939 &P\'eroux  \\
...          &2000    &{\it WFPC F450W}    &4556        &26.58                     &5351  &Bergeron  \\
...          &3600    &{\it WFPC F702W}    &6919        &27.97                     &5351  &Bergeron  \\
...          &767     &{\it NICMOS F160W}  &1607        &24.25              &7329  &Malkan    \\
Q0454$-$220  &2164    &WFC3/UVIS F336W	   &3355	    &29.31             &15939 &P\'eroux  \\
...          &2564    &WFC3/UVIS F475W	   &4774	    &29.34             &15939 &P\'eroux  \\
...          &1200    &{\it WFPC F702W}    &6917        &26.02                     &5098  &Burbidge  \\
Q1110$+$0048 &1500    &WFC3/UVIS F336W	   &3355	    &29.56             &15939 &P\'eroux  \\
...          &1200    &WFC3/UVIS F475W	   &4774	    &28.92             &15939 &P\'eroux  \\
...          &1200    &WFC3/UVIS F814W	   &8033	    &27.45             &15939 &P\'eroux  \\
J1130$-$1449 &2152    &WFC3/UVIS F336W	   &3355	    &29.42             &15939 &P\'eroux  \\	
...          &2552    &WFC3/UVIS F438W	   &4326        &29.08             &15939 &P\'eroux  \\	
...          &22000   &{\it WFPC F814W}    &8012        &28.65                     &9173  &Bechtold  \\
...          &2120    &{\it WFC3/IR F140W} &1392        &28.02              &14594 &Bielby    \\                
J1211$+$1030 &2224    &WFC3/UVIS F336W	   &3354	    &28.47             &15939 &P\'eroux  \\
...          &2000    &{\it WFPC F450W}    &4556        &28.21                     &5351  &Bergeron  \\                
...          &3600    &{\it WFPC F702W}    &6919        &27.43                     &5351  &Bergeron  \\                
Q1229$-$021  &2396    &WFC3/UVIS F336W	   &3355	    &28.67             &15939 &P\'eroux  \\
...          &2000    &{\it WFPC F450W}    &4556        &26.93                     &5351  &Bergeron  \\                
...          &4800    &{\it WFPC F702W}    &6919        &27.55                     &5351  &Bergeron  \\                
Q1342$-$0035 &2106    &WFC3/UVIS F336W	   &3355	    &28.54             &15939 &P\'eroux  \\
...          &2489    &WFC3/UVIS F475W	   &4774	    &29.20             &15939 &P\'eroux  \\
...          &2489    &WFC3/UVIS F814W	   &8033	    &28.12             &15939 &P\'eroux  \\
Q1345$-$0023 &500     &WFC3/UVIS F336W	   &3355	    &28.68             &15939 &P\'eroux  \\
...          &600     &WFC3/UVIS F475W	   &4774	    &29.56             &15939 &P\'eroux  \\
...          &1200    &WFC3/UVIS F814W	   &8033	    &27.80             &15939 &P\'eroux  \\
Q1431$-$0050 &2106    &WFC3/UVIS F336W	   &3355	    &29.06             &15939 &P\'eroux  \\
...          &2489    &WFC3/UVIS F475W	   &4774	    &28.79             &15939 &P\'eroux  \\
...          &2489    &WFC3/UVIS F814W	   &8033	    &28.08             &15939 &P\'eroux  \\
Q1515$+$0410 &2106    &WFC3/UVIS F336W	   &3355	    &29.14             &15939 &P\'eroux  \\
...          &2489    &WFC3/UVIS F475W	   &4774	    &29.62             &15939 &P\'eroux  \\
...          &2489    &WFC3/UVIS F814W	   &8032	    &26.82             &15939 &P\'eroux  \\
Q1554$-$203  &2411    &WFC3/UVIS F475W	   &4774	    &29.87             &15939 &P\'eroux  \\
...          &2576    &WFC3/UVIS F625W	   &6242	    &28.89             &15939 &P\'eroux  \\
...          &2176    &WFC3/IR F105W	   &10557	    &28.64             &15939 &P\'eroux  \\
J2131$-$1207 &2156    &WFC3/UVIS F336W	   &3355	    &29.07             &15939 &P\'eroux  \\
...          &2556    &WFC3/UVIS F475W	   &4774	    &29.26             &15939 &P\'eroux  \\
...          &1800    &{\it WFPC F702W}    &6919        &27.22                     &5143  &Macchetto \\                
Q2353$-$0028 &2106    &WFC3/UVIS F336W	   &3355	    &29.48             &15939 &P\'eroux  \\
...          &2489    &WFC3/UVIS F475W	   &4774	    &29.45             &15939 &P\'eroux  \\
...          &2489    &WFC3/UVIS F814W	   &8033	    &27.78             &15939 &P\'eroux  \\
\hline\hline 				       		         	 
\label{tab:JoO_HST}                                         
\end{tabular}			       			 	 
\end{center}			       			 	 
\end{table*}

\section{MUSE-ALMA Haloes Data Processing}
\label{sec:redu}

\subsection{Optical Integral Field Spectrograph VLT/MUSE Cubes}

\begin{figure}
	\includegraphics[width=8.5cm]{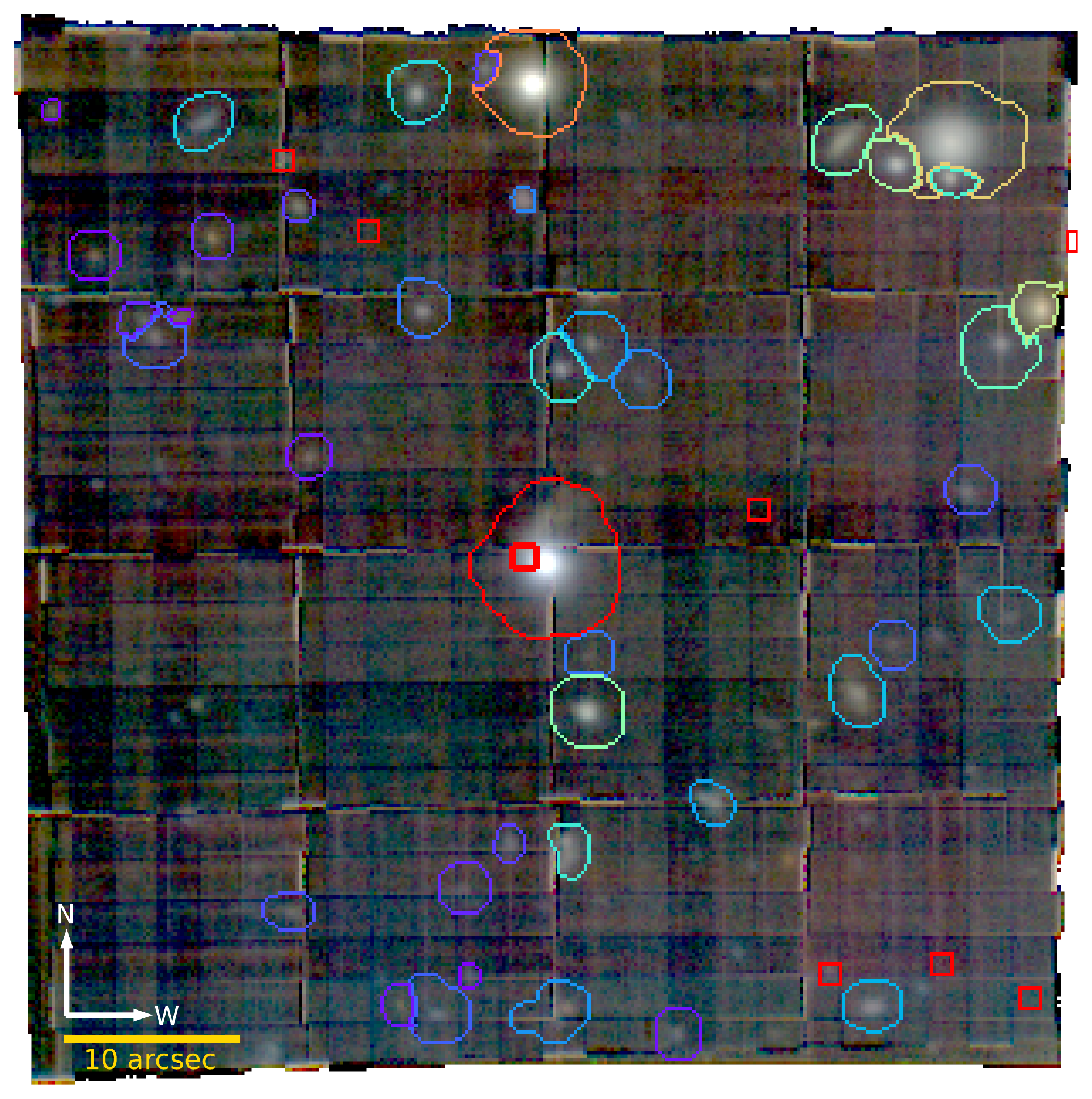}
    \caption{{\bf Example VLT/MUSE reconstructed RGB-colour image.} Reconstructed AO-assisted image of the combined exposures of the field of Q1110$+$0048 collapsed along the wavelength axis. 
    The total field-of-view is 59.9×60 arcsec$^2$. The total exposure time t$_{\rm exp}$=1410 s$\times$2 and the resulting image quality of the combined datacube is FWHM=0.52 arcsec measured at 7000 \AA. The quasar, with $z_{\rm quasar}$=0.76, is at the centre of the image. The objects visible in the field are bright in continuum. The colour contours illustrate of the automatically-generated segmentation map in the Johnson R-band of the same field ordered in terms of increasing flux. Brighter objects are warmer colours (red/orange) whereas fainter objects are blue/violet. Such segments are used to estimate the magnitudes of the sources in multiple bands with the \textsc{ProFound} R Package. The objects marked by small red squares are not continuum-detected but identified through their emission lines with the \textsc{mpdaf/MUSELET} package. The red square underneath the quasar PSF is also detected in HST observations (see Fig.~\ref{fig:HST_ima}). VLT/MUSE observations additionally provide spectroscopic and kinematic information for the majority of the objects in the field. }
    \label{fig:MUSE_ima}
\end{figure}

The data were reduced with the ESO MUSE pipeline \citep{Weilbacher15} and additional external routines for 
sky subtraction and extraction of the 1D spectra. Master bias, flat field images and arc lamp exposures based on data taken closest in time to the science frames were used to correct each raw cube. We checked that the flat-fields are the closest possible to the science observations in terms of ambient temperature to minimise spatial shifts. In all cases, we found the temperature difference to be below the canonical 0.5 degrees set to be the acceptable limit. Bias and flat-field correction are part of the ESO pipeline. The raw science data were then processed with the $scibasic$ and $scipost$ recipes. During this step, the wavelength calibration was corrected to a heliocentric reference. We note that MUSE operates in air, not vacuum. We checked the wavelength solution using the known wavelengths of the night-sky [\ion{O}{i}] and OH lines across the wavelength coverage of MUSE. We find a median discrepancy of $25$ km/s in the wavelength solution across the fields. The individual exposures were registered using the point sources in the field within the $exp\_align$ recipe, ensuring accurate relative astrometry. The astrometry of HST/WFC3 is checked using the central quasar as reference, and it is found to be accurate within sub-arcsec. Finally, the individual exposures were combined into a single data cube using the $exp\_combine$ recipe. The image quality of the final combined data is measured from a gaussian fit of the quasar at 7000\AA\ in the data cube. The resulting PSF FWHM values are listed in Table~\ref{tab:JoO_MUSE}. 

{To estimate flux errors, we measured fluxes in synthetic VLT/MUSE broad-band images created by applying HST filter curves\footnote{\url{http://svo2.cab.inta-csic.es/theory/fps/}}, and then compared these measured pseudo magnitudes to the actual HST data. While we refrain from systematically correcting the MUSE flux levels in order to keep information on the associated uncertainties, on four occasions (Q0138$-$0005, Q0152$+$0023, Q1431$-$0050 and {Q1515$+$0410}) where the flux differences were large ($>$50~per~cent) we adjusted the MUSE flux levels so as to match the HST values. In addition, for fainter objects ($> 23$ mag), we applied an additional error correction due to the volatility in the MUSE magnitudes \citep{Roth2018}. The standard deviation of the magnitude difference between MUSE and HST fluxes is added in quadrature to the error term returned by \textsc{ProFound} to obtain the final magnitude error. For the field without optical or near-IR HST data, we compared the measured fluxes of the central quasar and bright stars from pseudo Cousins V, Johnson R and SDSS \textit{r} and \textit{i} images with literature values. Overall, we estimate the uncertainties to be on average $\pm 30$~per~cent.} Finally, we checked the VLT/MUSE absolute astrometry using the HST/WFC3 data when available. We applied appropriate systemic offsets of the order 1 arcsec to the VLT/MUSE cubes.

The removal of OH emission lines from the night sky is accomplished with additional purpose-developed codes. The $scipost$ recipe is first performed with sky-removal method "simple" on, which directly subtracts a sky spectrum created from the data, without regard to the line spread function (LSF) variations. After selecting sky regions in the field, we create Principal Component Analysis (PCA) components from the spectra which are further applied to the science datacube to remove sky line residuals \citep{Husemann16, peroux2017}. This method is required as an addition to the ESO pipeline to significantly improve the sky subtraction over large parts of the MUSE field-of-view. The RGB-colour image of an example field is shown in the left panel of Fig.~\ref{fig:MUSE_ima}.

\subsection{ALMA mm Cubes}

We describe here the steps taken to reduce the ALMA data. Additional details can be found in \cite{peroux2019, Szakacs21, Klitsch21} for the PI data and \cite{Klitsch18} for ALMACAL. We started the data reduction with the pipeline-calibrated $u,v$ data as delivered by ALMA. Additional data reduction steps were carried out with the Common Astronomy Software Applications (\textsc{CASA}) software package. Minor manual flags were added to remove some $u,v$ data with strongly outlying amplitudes. Some of the quasar in the centre of the science fields are very bright at mm frequencies ($\ge$100 mJy), which makes them ideal sources for self-calibration, in both amplitude and phase.

Self-calibration was carried out by firstly using \textsc{tclean} to Fourier transform the $u,v$ data into a continuum image and deconvolving. Self-calibration was then performed using the \textsc{gaincal} and \textsc{applycal} on individual measurement sets to produce corrected $u,v$ data, after which the data were re-imaged to create an improved continuum map. We applied one round of phase self-calibration and one round of amplitude and phase self-calibration. The next step was to subtract the bright continuum source from the field using the \textsc{uvsub}. We then created a cube with \textsc{tclean}, setting the pixel size so as to oversample the beam sufficiently and using a `robust' weighting scheme with a Briggs parameter of 0.5 or 1. Residual continuum signatures around the imperfectly subtracted quasar were removed using \textsc{uvcontsub}. After searching for emission lines, we corrected the cube for the primary beam using \textsc{pbcor} in order to correctly measure their flux. We note that the resulting FWHM of the primary beam of ALMA in Band 3 is $\sim$62\arcsec, conveniently matching the VLT/MUSE field-of-view. We refer the reader to \cite{Klitsch18, peroux2019, Klitsch21, Szakacs21} for examples of the resulting ALMA datacubes.

\subsection{HST Broad-Band Imaging}

\begin{figure*}
	\includegraphics[width=18cm]{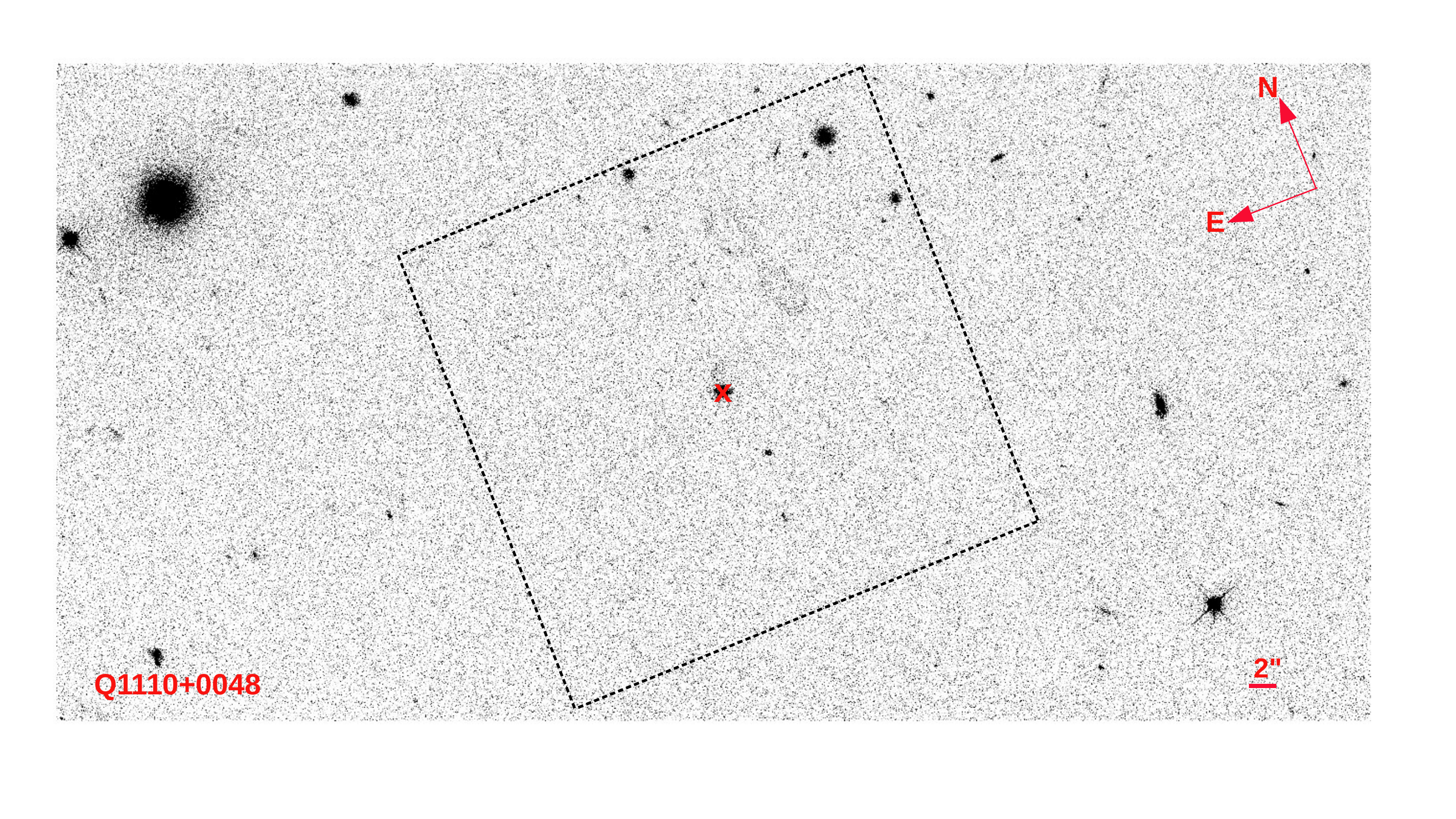}
	\includegraphics[width=16cm]{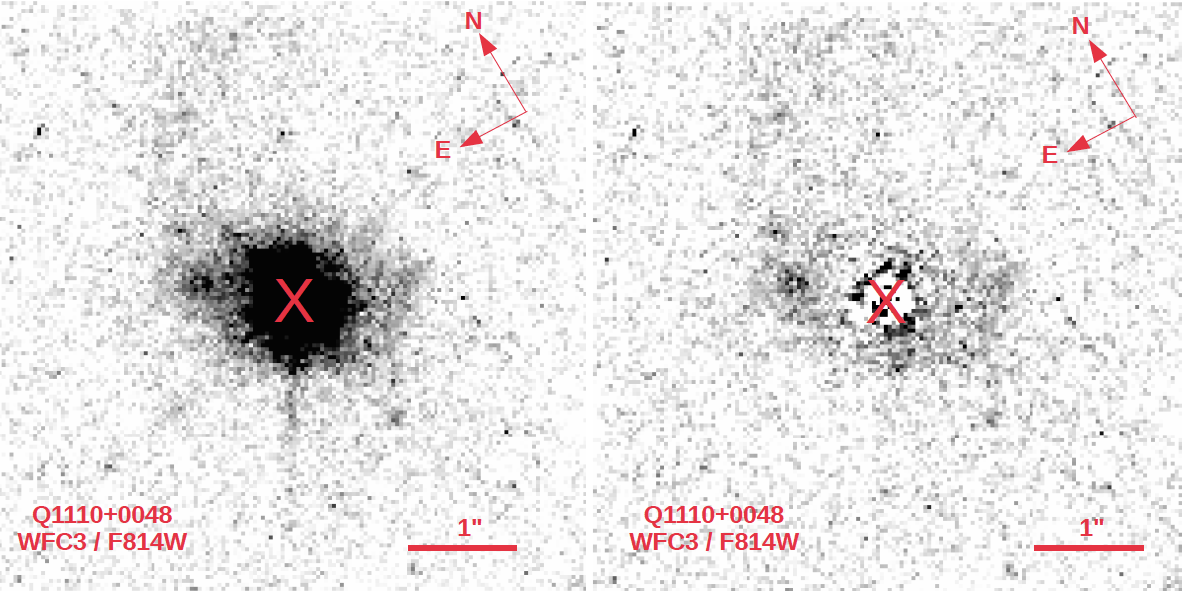}
    \caption{{\bf Example HST broad-band imaging.} Reconstructed image of the combined exposures of the field of Q1110$+$0048 in the F814W UVIS filter-band. The total exposure time t$_{\rm exp}$ is 1200 sec. The quasar, with $z_{\rm quasar}$=0.76, is at the centre of the images and is marked with a red cross. {\it Top panel:} The full F814W UVIS broad-band image. 
    {The MUSE field-of-view is overlaid as a dotted square.}
    The image shows multiple objects which are bright in continuum. {\it Bottom left panel:} Zoom-in of the same image.
    The bright quasar displays strong diffraction spikes. {\it Bottom right panel:}  Same zoom-in image after a careful PSF-subtraction has been performed to remove the quasar image as described in the text.  The image clearly shows an additional object north-east of the quasar, which is also detected in emission in the VLT/MUSE cube (see Fig.~\ref{fig:MUSE_ima}).    }
    \label{fig:HST_ima}
\end{figure*}

\subsubsection{Data Reduction}

The WFC3 data were reduced with the \textsc{calwf3} pipeline. The pipeline processing steps include bias subtraction, dark subtraction, and flat fielding. Each individual reduced exposure was multiplied by the pixel area map provided on the HST/WFC3 photometry website in order to perform flux calibration. Bad pixels, saturated pixels, and pixels affected by cosmic rays  were masked using the data quality file provided with each science frame. Subpixel grids were constructed on the individual exposures for the purpose of achieving accurate alignment of dithered images, using grids of 5$\times$5 pixels for the full images, and 10$\times$10 pixels for the central portions of the images zoomed in on the quasar. The individual sub-pixeled images were sky-subtracted and then shifted with respect to each other as needed to align them. In some cases, the individual exposures were taken at different roll angles, resulting in relative rotation of the field. In such cases, the tasks \textsc{Tweakreg} and \textsc{Astrodrizzle} were used to rotate the images before alignment. The sky-subtracted, aligned individual exposures were then median-stacked to produce the final science images. The upper panel of Fig.~\ref{fig:HST_ima} shows an example WFC3 UVIS image for one of the fields in the filter F814W.

The archival WFPC data were reduced with the \textsc{calwp2} pipeline, which includes the steps of bias subtraction, dark subtraction, and flat field correction. The data quality file provided with each science data file was used to flag bad pixels and saturated pixels. The \textsc{IRAF} task \textsc{CRREJ} was used to correct pixels hit by cosmic rays in each science frame. 
The resulting images were further processed using identical steps as described above for WFC3 data (subpixeling and sky-subtraction of the individual dithered exposures, alignment of the individual images including rotation with \textsc{Tweakreg} and \textsc{Astrodrizzle} if necessary, and median stacking of the aligned images). We note that in the two cases where rotational alignment using \textsc{Tweakreg} and \textsc{Astrodrizzle} was performed, the multiplication by pixel area map was not performed, since the varied pixel area is fixed by the drizzle process. In all other cases, the multiplication by the pixel area map was necessary since the FLC images were used directly. 


\subsubsection{Subtraction of Quasar Point Spread Function}

The primary difficulties in detecting in emission the galaxy or galaxies producing a quasar absorption system are the faintness of the galaxy relative to the background quasar and the small angular separation between the galaxy and the quasar. To make it possible to detect the galaxy's continuum emission, it is essential to remove the contamination from the quasar by subtracting the point spread function (PSF) from the quasar image. A major advantage of HST, in this context, over ground-based imaging systems is that the PSF of HST cameras is far better defined and stable compared to the PSFs of ground-based imaging systems (even those using adaptive optics). Of course, given the diffraction-limited nature of HST imaging, the PSF depends on the wavelength of observation, and is thus different for different filters. Moreover, the PSF can vary spatially across the field of view. Guided by our past experience with PSF subtraction for detecting galaxies in quasar fields \citep[e.g.][]{kulkarni2000, kulkarni2001, Chun10, Straka11, augustin2018}, we constructed the PSF using our own observations, rather than relying on theoretical PSF models. For constructing the PSF in each filter for each quasar field, we used observations of all other quasar fields from our sample in that same filter, combining all such observations after masking objects other than the central quasars, subtracting the sky from each image, aligning the different images spatially so that the diffraction spikes overlap, and scaling them as needed in flux to make the flux levels match in the outer wings of the PSF. The such reconstructed PSF  was then aligned and matched in flux with the studied quasar and subtracted from the quasar to reveal any underlying galaxies. This strategy proved successful, and led to robust removal of diffraction spikes, enabling detections of faint galaxies previously hidden under the quasar PSF in several of our fields. A similar approach was adopted while performing the PSF subtraction for the archival WFC3 or WFPC images, constructing the PSF from either an isolated star in the field or from a combination of different quasar images in the same filter. The lower two panels of Fig.~\ref{fig:HST_ima} demonstrate the effectiveness of our PSF subtraction strategy. While the residual flux level close to the subtracted quasar's centre (marked with an ``X") is not always zero, it is possible to see faint extended objects which were hidden under the PSF a little further away from the quasar centre. 

\section{Searching for Extra-galactic Objects}
\label{sec:analysis}

The identification of galaxies in our data is necessary to explore the cosmic baryon cycle.
We used the VLT/MUSE observations of the quasar fields, together with the HST wide-field broad-band imaging to search for all extra-galactic objects in each field. Several analysis techniques were applied to maximize the completeness of the search of different types of objects (star forming-galaxies, passive galaxies, faint objects etc.). VLT/MUSE datacubes can be viewed as individual narrow-band (NB) images at each wavelength slice or a combined continuum image over the whole observed wavelength range, allowing for two types of object searches in the data: a single spectral-line search and identification of the sources seen in continuum. 

\subsection{Detecting Emitting Galaxies in VLT/MUSE Cubes}

We expect the presence in the VLT/MUSE data of objects with emission lines but no detectable continuum.
Some of these sources may only exhibit a single emission line and the interpretation of these sources requires further inspection. We used the MUSE Line Emission Tracker (\textsc{MUSELET}) module of the MPDAF\footnote{\url{mpdaf.readthedocs.io}} package \citep{Piqueras17} to systemically search for emission-line objects. 
\textsc{MUSELET} creates synthetic narrow-band images (width of $7.25$ \AA) at each wavelength plane of a cube that are then passed through \textsc{SExtractor} to find objects with emission lines.
Regions near the night sky emission lines at $5577$ and $6300$ \AA\ were excluded to limit the number of false detections caused by residues from the sky subtraction.
The list of objects from MUSELET will overlap with the continuum objects detected, and these were removed to prevent duplicates.

In some cases, we detected galaxies close to the quasar position (within 1 arcsec), blended with the quasar PSF. In order to uncover such object, we performed a careful spectral PSF subtraction within \textsc{QFitsView}. To this end, the quasar PSF was fitted as a function of wavelength so that it became possible to detect faint emitting galaxies despite the bright quasar contribution. This technique however provides limited information on the continuum properties of such galaxy \citep[although see][]{rupke17,helton21}.

\subsection{Detecting Continuum Galaxies in VLT/MUSE Cubes}

The \textsc{ProFound} R Package\footnote{\url{https://github.com/asgr/ProFound}} \citep{Robotham18} was used to detect continuum sources in the MUSE fields.
First, white-light and synthetic Cousins V, Johnson R and SDSS \textit{r} and \textit{i} band images were created using MPDAF.
\textsc{ProFound} was initially used on the white-light image to produce a preliminary segmentation map.
Due to background residuals in some fields, the $profoundSegimFix$ function was used to manually modify affected segments and remove false detections \citep{Bellstedt20, Foster21}. An example segmentation map is shown in Fig.~\ref{fig:MUSE_ima} for illustration.
Finally, \textsc{ProFound} was run in multi-band mode to determine the photometric properties in the V, R, \textit{r} and \textit{i} bands by collapsing the MUSE cubes in wavelength ranges specific to different bands.

The choice of \textsc{ProFound} over \textsc{SExtractor}\footnote{\url{https://www.astromatic.net/software/sextractor}} stems from the difference in nature between the VLT/MUSE white-light images and HST broad-band images. 
While objects in the HST images are typically separated due to the lower PSF FWHM, the blending of adjacent objects occurs in our VLT/MUSE fields. 
This increases the importance of obtaining accurate segmentation maps by ensuring apertures do not erroneously combine segments, which was seen to occur more often in \textsc{SExtractor} \citep{Robotham16, Robotham18}.
Both programs were used to search for continuum sources and it was found that \textsc{ProFound} produced more accurate segmentation maps.
For bright and isolated point sources, the photometric properties measured using both alogrithms were found to be consistent within $5$~per~cent.

Last, we also used HST detection of faint continuum sources as prior information for assessing the VLT/MUSE possible sources. The sky position of the HST objects were used as priors to extract spectra from the VLT/MUSE cubes which were then processed in the same way as any other continuum detected objects. This process led to a couple of additional detections.


\subsection{Detecting Continuum Galaxies in HST Broad-Band Imaging}

The \textsc{Astropy} package \textsc{Photutils} was used on the processed HST 
broad-band images in each filter for each field to search for all objects and to perform 
photometry of the detected objects. Apertures were optimised by repeating the photometry 
on elliptical apertures of increasing sizes for each object and adopting the aperture 
size beyond which the flux of that object stayed constant (taking care to check that the 
aperture did not include multiple objects). Because this was done separately {for each band}, we also checked the aperture corrections. To this end, we estimated the 
Sersic index and {the effective} radius of the {objects using GALFIT. We next 
computed the theoretical light profile for the Sersic index and the effective radius thus determined. Using this profile, we examined the flux enclosed within apertures of radius $r$ (i.e., flux integrated from 0 to $r$) as a function of $r$.} We then determined the minimum radius above which {the enclosed} flux remains constant {(i.e., stops increasing)}. We find that this radius is less {than} the one used in both \textsc{SExtractor} and \textsc{ProFound}, indicating that the aperture corrections are negligible and the  {magnitudes of the aperture corrections are comparable for the different bands.} The package \textsc{SExtractor} was used to classify each object detected in the field as a star or a galaxy {using the CLASS$\_$STAR parameter in SExtractor}. As an additional check on the photometry, fluxes 
computed using \textsc{SExtractor} were compared with those obtained using 
\textsc{Photutils}, and were found to agree closely (within $\sim$2-5~per~cent). Once the 
photometry in each filter was completed, cross-matching was performed between the objects 
detected in the different filters using the tool \textsc{Topcat} \citep{Taylor05}, to 
construct a catalog of all detected objects in each field, including the photometry for 
each object in all filters. In cases where an object detected in one filter was not detected in another filter, a 3-$\sigma$ magnitude limit was calculated for the filter with the non-detection by measuring the 3-$\sigma$ noise level (over the number of pixels occupied by the object in the filter with the detection). Finally, the objects in the catalog were examined visually to identify any spurious detections near the edges of the images (caused by artifacts), and such false objects were removed from the catalog.

\section{Galaxies' Physical Properties}
\label{sec:cat}

\subsection{Optical Spectral Extraction}

\begin{figure*}
	\includegraphics[width=10.5cm]{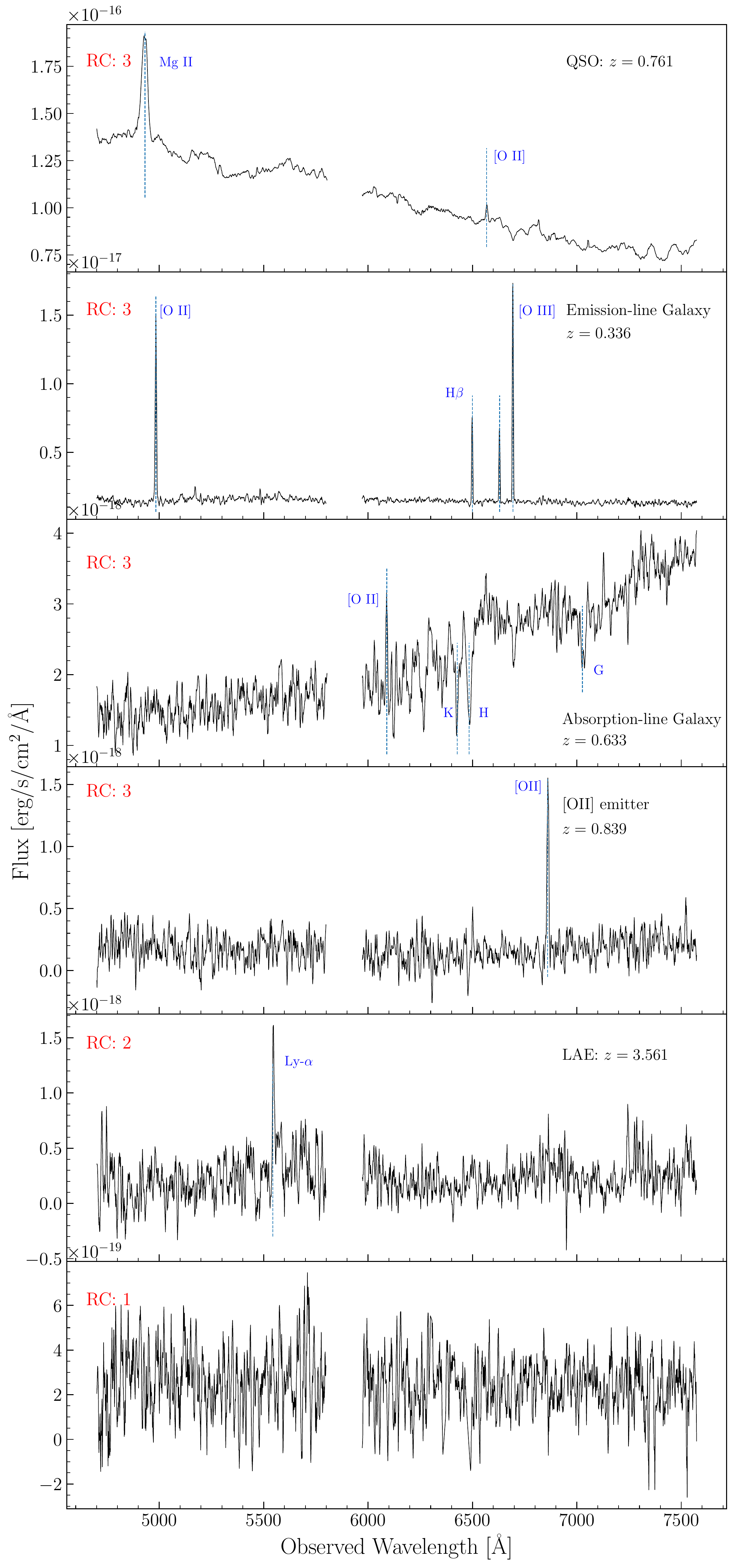}
    \caption{{\bf Example VLT/MUSE spectra.} Five-pixel boxcar/moving-average spectra of six objects observed as part of the Q1110$+$0048 VLT/MUSE cube. 
    {The assigned spectral redshift confidence ("RC") is given in red at the top left of each panel.
    The [\ion{O}{ii}] emitter at $z = 0.839$ in the fourth panel from the top has its doublet resolved in the unsmoothed spectrum and hence, it was graded as a 'redshift confidence: 3'.}
    The wavelengths not plotted correspond to the notch filter between 5820-5970 \AA\ inherent to AO-assisted observations. 
    The spectral resolution is R=1770 at 4800 \AA\ and R=3590 at 9300 \AA\ resampled to a spectral sampling of 1.25 \AA/pixel.    }
    \label{fig:MUSE_spec}
\end{figure*}

The spectral extraction method differs for the continuum objects detected using \textsc{ProFound} and emission-line objects found by MUSELET. 
Due to the large variability in seeing conditions between the fields, the ideal aperture size for extracting the spectra of our continuum sources will differ significantly across cubes.
Additionally, objects within cubes have varying sizes and morphologies.
To obtain the optimal spectra for {for redshift determination} in each field, a range of circular apertures with sizes ranging from $0.3$ to $2.0$ arcsec in radius was used.
{The lower limit is set by the PSF FWHM of the MUSE data cube and all the apertures have a minimum diameter above this value.}
For each aperture radius, the signal-to-noise (S/N) of the spectrum at the wavelength interval spanning $\sim 5000 - 5200$ \AA \  was calculated.
This wavelength region is chosen as there are no night sky emission lines or artefacts in this range, and the aperture size where S/N is maximised determined our final extracted spectrum.
Redder wavelength planes ($7050 - 7200$ and $8100 -8250$ \AA) were also tested and the change in S/N consistently varied with aperture size across the different wavelengths.
To prevent our apertures from including neighbouring objects, regions outside the segment containing the object are masked.
While the apertures used are circular, masking based on the segmentation map effectively changes the aperture shape to the shape of the segment and by extension, our object. {While these are not the spectra used for emission lines flux measurement, we note that $> 95$\% of an object's total flux is captured using this method to extract spectra.}

For the emitters detected by MUSELET and sources extracted using the broad-band HST images as a prior, a circular aperture of radius $0.5$ arcsec is used to obtain spectra.
These objects are not expected to be detected in continuum and hence the choice of aperture size becomes less relevant. Example VLT/MUSE spectra are displayed in Fig.~\ref{fig:MUSE_spec}.


\subsection{Spectral Classification}

The extracted VLT/MUSE spectra are used to both identify the sources and to estimate the redshift of extra-galactic objects. To this end, we use the MARZ tool \citep{Hinton16} with the M. Fossati fork\footnote{\url{matteofox.github.io/Marz/}}. This fork includes additional high-redshift templates and high-resolution templates well suited for VLT/MUSE data. 
MARZ provides both a visualisation tool and a template cross-correlation tool for each source with quasar, galaxy and stellar template spectra. The results are visually inspected by two experts (SW and CP) to confirm the nature and redshift of the source based on continuum level and shape as well as on detected emission and absorption lines. Faint sources 
can have spectra of insufficient quality to attempt a redshift determination using cross-correlations, though some have bright 
emission lines and are included in the catalogue following the search for emission lines described earlier. {The redshift success rates ranges from 100\% at $r_{\rm mag} = 20$ down to typically 60\% at $r_{\rm mag} \sim 25$.}

We therefore provide {redshift confidences} for each source. They are as follows: (1) spectra without emission or absorption features and no redshift estimate is possible, (2) a redshift measure is possible as the spectra contains low S/N emission and absorption lines, and (3) a {robust redshift is determined} where there is a high S/N emission line that is clearly {resolved [OII] doublet} or {asymmetric } Lyman-$\alpha$, one high S/N emission line with other fainter absorption or emission features, or with multiple clear emission and/or absorption lines, and finally (6) stellar objects. These results are also part of the tables made available, as summarised by the column entries presented in Table~\ref{tab:mag}.


\subsection{Matched Catalogues}

In order to facilitate the access to the information reported in this
work, we have assembled master tables for each targeted field with all the information
derived for the emitting galaxies from the VLT/MUSE and HST detections. To match various HST filters and additional HST imaging with VLT/MUSE results, we use the tool \textsc{TopCat} to produce these master tables. There are three distinct types of entries: i) objects detected in MUSE cubes but not in HST images {(typically emission line objects with faint continuum)}, ii) objects detected in HST images but not in MUSE cubes, and iii) objects detect in HST images but outside the MUSE field-of-view. Objects detected in MUSE are listed first, followed by objects only detected in HST imaging. The id order is in descending order according to the
object flux in the reddest filter. The tables list a unique id number, sky coordinates, 
\textsc{SExtractor}-based star/galaxy classification parameter, multi-wavelength photometry, and spectroscopic redshifts with associated flag when available and finally the source detection method. The number "999" means there is no information in this entry: i) either there are no such HST filter observed, or ii) the object is not detected in HST and as such as no "object classifier", or iii) the redshift could not be estimated. An "999" entry in the error columns indicates that the corresponding measure is an upper limit from a non-detection. The master-tables are available as machine-readable on-line material. Table~\ref{tab:mag} summarises the column entries.

\begin{table*}
\begin{center}
\caption{  {\bf MUSE-ALMA Haloes galaxy properties.} 
  The table lists the continuum and spectra-based measurements of all extragalactic objects in the 19 fields observed with VLT/MUSE, {ALMA} and HST. The table includes the object sky position, the HST and VLT/MUSE-band photometry and spectroscopic redshifts {and CO J-transitions covered by ALMA} when available. {CO J-transitions from CO(1-0) up to CO(10-9) are searched for. We additionally check whether the [\ion{C}{ii}] transition at 158 $\mu$m is covered.
  Absolute magnitudes and rest-frame colours are calculated using two HST filters. The symbols "XXX" refer to header types varying from field to field.}
  There are three distinct types of entries: i) objects detected in MUSE cubes but not in HST images, ii) objects detected in HST images but not in MUSE cubes, and iii) objects detect in HST images but outside the MUSE field-of-view. The number "999" means there is no information in this entry: i) either there are no such HST filter observed, or ii) the object is not detected in HST and as such as no "object classifier", or iii) the redshift could not be estimated. An "999" entry in the error columns indicates that the corresponding measure is an upper limit from a non-detection. This extract shows the column entries while the full master-tables are available for every field as machine-readable on-line material.
}
\begin{tabular}{clcl}
\hline\hline
Column &Name &Format &Description\\
\hline
1  &id                              &INTEGER       &Object identification number                                                                 \\
2  &RA                              &FLOAT        &Right Ascension in decimal degrees (J2000)                                                   \\                 
3  &Dec                             &FLOAT        &Declination in decimal degrees (J2000)                                                       \\
4  &Object classifier               &FLOAT      &\textsc{SExtractor} star/galaxy classification parameter ($<$0.95 likely extra-galactic)        \\
5  &F336W                           &FLOAT        &WFC3/UVIS F336W magnitude                                                                    \\
6  &F336W$_{\rm err}$                &FLOAT        &WFC3/UVIS F336W magnitude error                                                               \\
7  &F438W                           &FLOAT        &WFC3/UVIS F438W magnitude                                                                    \\  
8  &F438W$_{\rm err}$                &FLOAT        &WFC3/UVIS F438W magnitude error                                                               \\
9  &F450W                           &FLOAT        &WFC3/UVIS F450W magnitude                                                                    \\  
10 &F450W$_{\rm err}$                &FLOAT        &WFC3/UVIS F450W magnitude error                                                               \\
11 &F475W                           &FLOAT        &WFC3/UVIS F475W magnitude                                                                    \\  
12 &F475W$_{\rm err}$                &FLOAT        &WFC3/UVIS F475W magnitude error                                                               \\
13 &F625W                           &FLOAT        &WFC3/UVIS F625W magnitude                                                                    \\  
14 &F625W$_{\rm err}$                &FLOAT        &WFC3/UVIS F625W magnitude error                                                               \\
15 &F702W                           &FLOAT        &WFPC F702W magnitude                                                                           \\  
16 &F702W$_{\rm err}$                &FLOAT        &WFPC F702W magnitude error                                                                    \\
17 &F814W                           &FLOAT        &WFC3/UVIS F814W magnitude                                                                    \\  
18 &F814W$_{\rm err}$                &FLOAT        &WFC3/UVIS F814W magnitude error                                                               \\
19 &F105W                           &FLOAT        &WFC3/IR F105W magnitude                                                                    \\  
20 &F105W$_{\rm err}$                &FLOAT        &WFC3/IR F105W magnitude error                                                               \\
21 &F140W                           &FLOAT        &WFC3/IR F140W magnitude                                                                    \\  
22 &F140W$_{\rm err}$                &FLOAT        &WFC3/IR F140W magnitude error                                                               \\
23 &NicmosF160W                     &FLOAT        &Nicmos F160W magnitude                                                                      \\
24 &NicmosF160W$_{\rm err}$          &FLOAT        &Nicmos F160W magnitude error                                                                \\
25 &V$_{\rm mag}$                    &FLOAT        &MUSE Cousins V-band magnitude                                                               \\
26 &V$_{\rm err}$                    &FLOAT        &MUSE Cousins V-band magnitude error                                                          \\
27 &R$_{\rm mag}$                    &FLOAT        &MUSE Johnson R-band magnitude                                                               \\
28 &R$_{\rm err}$                    &FLOAT        &MUSE Johnson R-band magnitude error                                                         \\
29 &r$_{\rm mag}$                    &FLOAT        &MUSE Sloan \textit{r}-band magnitude                                                                   \\
30 &r$_{\rm err}$                    &FLOAT        &MUSE Sloan \textit{r}-band magnitude error                                                             \\
31 &i$_{\rm mag}$                    &FLOAT        &MUSE Sloan \textit{i}-band magnitude                                                                   \\
32 &i$_{\rm err}$                    &FLOAT        &MUSE Sloan \textit{i}-band magnitude error                                                             \\
33 &Redshift                        &FLOAT        &Redshift from \textsc{MARZ}+visual inspection                                                                     \\
34 &Redshift {Confidence}             &INTEGER          &{Confidence} of the redshift                                                                     \\
35 &origin                          &STRING       &Source detection method                                                                      \\
{36} &FXXXWAbs    &FLOAT       &Absolute magnitude in a HST filter given by the header                                            \\
{37} &FXXXWAbs$_{\rm err}$    &FLOAT       &Absolute magnitude error                                                                     \\
{38} &FXXXW-FXXXW$_{\rm col}$    &FLOAT       &Rest-frame colour using filters given in the header                                            \\
{39} &(FXXXW-FXXXW)$_{\rm err}$     &FLOAT       &Rest-frame colour error                                                                     \\
{40+} & CO(X-X)$_{\rm freq}$     &STRING       &Observed frequency of a given CO J-transition in Hz                                                  \\
{41+} & CO(X-X)$_{\rm band}$ &STRING       &ALMA band covering the given CO J-transition
  \\
{42+} & CO(X-X)$_{\rm exp}$     &FLOAT        &Sum of the exposure times covering the given CO J-transition                                    \\
\hline\hline 				       			 	 
\label{tab:mag}
\end{tabular}			       			 	 
\begin{minipage}{180mm}
\end{minipage}
\end{center}			       			 	 
\end{table*}

\section{Conclusions}
\label{sec:ccl}

This paper presents the MUSE-ALMA Haloes survey which is designed with the broad goals of quantifying the physics of
the multi-phase gas associated with the CGM regions of galaxies.
This program adds to a number of existing efforts which use VLT/MUSE IFS observations to build a comprehensive picture of gas flows in galaxies. Notably, the Muse Gas Flow and Wind (MEGAFLOW, PI: N. Bouch\'e) survey has focused on MgII metal absorbers at intermediate redshift. The particular focus of these studies include quantifying the physical properties of outflows \citep{schroetter2016, schroetter2019, zabl2019} and inflows \citep{zabl2019}. Based on IRAM/NOEMA results, the molecular gas content of these objects appears to be low so as to require deep mm-observations \citep{Freundlich21}. A second program part of the MUSE intrument building team Guaranteed Time Observations (GTO), is the MUSEQuBES survey (PI: J. Schaye). While MUSEQuBES comprises a low and high-redshift component, early results concentrate on the Ly$\alpha$ emitters at $z>$3. The results show that the wind velocities correlate with the circular velocities, with indications of stronger winds for more massive galaxies  \citep{Muzahid20}. In addition, the findings of the survey suggest that Ly-$\alpha$ emitters surrounded by more neutral gas tend to have higher star formation rates \citep{muzahid2021}. Finally, the MUSE Analysis of Gas around Galaxies (MAGG) is an open-access Large Program effort (PI: M. Fumagalli). While MAGG initially concentrates on strong \hi-absorbers at $z>$3 \citep{lofthouse2020}, results also cover MgII absorbers at intermediate redshifts \citep{Dutta20, Dutta21} and gas properties of quasars themselves \citep{Fossati21}. In particular, the results indicated the vast majority of the MgII absorbers are associated with more than one galaxy \citep{Dutta20}. The findings show that environmental processes have a significant impact on the distribution of metals around galaxies as traced by MgII and CIV \citep{Dutta21}. One of the great successes of these IFS surveys is to solve a 2-decade long challenge 
by routinely identifying faint galaxies at the redshift of known quasar absorbers.

In this landscape, the MUSE-ALMA Haloes program provides a complementary perspective on strong \hi\ absorbers at $z<$0.85. Unique to the survey is the multi-wavelength approach which combines VLT/MUSE observations of 19 quasar fields with ALMA observations, offering a unique insight on the molecular gas content of these objects, and the multi band HST imaging to constrain the stellar population properties. Specifically, this paper describes the scientific motivation and background for the MUSE-ALMA Haloes survey, and the sample selection of the so-called primary absorbers and additional targets. We also report the design and execution of the multi-facilities programs made of an ensemble of VLT/MUSE, ALMA and HST observations based on PI and archival datasets. We describe the data processing of these various datasets and find that overall the performance of our survey is compatible with our initial goals. Importantly, this paper describes the global properties of the galaxy-selected sample targeted as part of the survey. This comprises an overview of the major components of the steps used to produce catalogues of magnitudes and redshift determination based VLT/MUSE spectroscopy.
We detail the estimate of the multi-band continuum magnitude measurements in multiple (VLT/MUSE and HST) bands and spectral extraction. We also report 703 redshift estimates from template matching coupled with a dedicated visual inspection.  
We present a matched catalogue including magnitudes of a total of 3658 extra-galactic sources and spectroscopic redshifts when available.






With the transformative capabilities of VLT/MUSE, ALMA and HST
working in concert, MUSE-ALMA Haloes advances our view of the multi-phase gas properties and association with the physical conditions of the stellar component of galaxies. A series of initial papers already presented exemplary case of such scientific
results of the MUSE-ALMA Haloes program \citep{peroux2017, Klitsch18, peroux2019, hamanowicz2020, Klitsch21, Szakacs21}. 
This multi-wavelength dataset also brings new information on other processes not showcased in this work, including e.g. AGN physics and high-redshift Ly$\alpha$ emitters. Together, these data will allow the identification and characterisation of the physical state of inflowing and outflowing gas, the fate of galactic winds (escaping or recycling) and the mass, molecular content and metallicity of the ejected material as a function of galaxy properties (redshift, SFR, metallicity, gas and stellar masses, morphology, environment). We emphasize that statistical approaches and large samples are required to be able to fully characterise the CGM and advance our understanding of the cosmic baryon cycle. The timely nature of this program can be appreciated with the imminent launch of JWST and the currently active preparation of ELT suits of instruments \citep{Ramsay21}.






\section*{Data Availability}

Data directly related to this publication and its figures are available upon request. The raw data can be downloaded from the public archives with the respective project codes.

\section*{Acknowledgements}


AK and VPK gratefully acknowledge partial support from a grant from the Space Telescope Science Institute for HST program 
GO-15939, and additional support from NASA  grant NNX17AJ26G (PI V. Kulkarni). RA was supported by HST GO \#15075. AK gratefully acknowledges support from the Independent Research Fund Denmark via grant number DFF 8021-00130. We thank ISSI (\url{https://www.issibern.ch/}) for financial support. We are grateful to the ESO Paranal and Garching, ALMA and Space Telescope Science Institute staff for performing the observations and the instrument teams for making reliable instruments. 
This paper makes use of the following ALMA data:
ADS/JAO.ALMA\#2016.1.01250.S, ADS/JAO.ALMA\#2017.1.00571.S and ADS/JAO.ALMA\#2018.1.01575.S. ALMA is a partnership of ESO (representing its member states), 
NSF (USA) and NINS (Japan), together with NRC (Canada), NSC and ASIAA (Taiwan), and KASI 
(Republic of Korea), in cooperation with the Republic of Chile. 
The Joint ALMA Observatory is operated by ESO, AUI/NRAO and NAOJ. The data presented in this paper were
obtained from the Mikulski Archive for Space Telescopes
(MAST). STScI is operated by the Association of Universities for Research in Astronomy, Inc., under NASA
contract NAS5-26555. Support for MAST for non-HST
data is provided by the NASA Office of Space Science
via grant NNX09AF08G and by other grants and contracts.


\appendix


\bibliographystyle{mnras}
\bibliography{UpdatedBiblio, NewBiblio}

\end{document}